\documentclass[useAMS,usenatbib]{mn2e} 

\usepackage{subfigure}
\usepackage{rotating,caption}
\usepackage{lscape}
\usepackage{amsmath}
\usepackage{natbib}
\usepackage{rotating}
\usepackage[bookmarks=true]{hyperref}

\usepackage{color}
\definecolor{darkgreen}{rgb}{0,0.5,0}
\definecolor{darkblue}{rgb}{0,0,0.55}
\hypersetup{colorlinks,linkcolor=blue,filecolor=darkgreen,citecolor=blue}

\newcommand{\msun}{$~M_{\odot}$}

\newcommand{\HII}{H~{\sc ii}}

\begin{document}

\title[NGC\,7538 : IRS\,1-3 and IRS\,9]
      {NGC\,7538 : Multiwavelength Study of Stellar Cluster Regions associated with IRS\,1--3 and IRS\,9 sources.} 

\author[K. K. Mallick et al.] 
       {K. K. Mallick,$^1$\thanks{E-mail: kshitiz@tifr.res.in} D. K. Ojha,$^1$ M. Tamura,$^2$ A. K. Pandey,$^3$ S. Dib,$^{4,5}$ 
        \newauthor 
        S. K. Ghosh,$^{1,6}$ K. Sunada,$^7$ I. Zinchenko,$^{8,9}$ L. Pirogov,$^{8,9}$ and M. Tsujimoto$^{10}$ \\
       $^1$ Department of Astronomy and Astrophysics, Tata Institute of Fundamental Research, Homi Bhabha Road, Colaba, \\
            Mumbai 400 005, India \\
       $^2$ National Astronomical Observatory of Japan, Mitaka, Tokyo 181-8588, Japan \\
       $^3$ Aryabhatta Research Institute of Observational Sciences, Manora Peak, Nainital 263 129, India \\
       $^4$ Niels Bohr International Academy, Niels Bohr Institute, Blegdamsvej 17, DK-2100, Copenhagen, Denmark \\
       $^5$ Centre for Star and Planet Formation, University of Copenhagen, {\O}ster Voldgade 5-7., DK-1350, Copenhagen, Denmark \\
       $^6$ National Centre for Radio Astrophysics, Tata Institute of Fundamental Research, Pune 411 007, India \\
       $^7$ Mizusawa VLBI Observatory, NAOJ, 2-12 Hoshi-ga-oka, Mizusawa-ku, Oshu-shi, Iwate 023-0861, Japan \\
       $^8$ Institute of Applied Physics, Russian Academy of Sciences, 46 Uljanov str., Nizhny Novgorod 603950, Russia \\
       $^9$ Lobachevsky State University of Nizhni Novgorod, 23 Prospekt Gagarina, 603950, Nizhni Novgorod, Russia \\
       $^{10}$ Japan Aerospace Exploration Agency, Institute of Space and Astronautical Science, 3-1-1 Yoshinodai, Chuo-ku, 
               Sagamihara, \\ Kanagawa 252-5210, Japan 
       }

\maketitle

\begin{abstract}

We present deep and high-resolution (FWHM $\sim$ 0\farcs4) near-infrared (NIR) 
imaging observations of the NGC\,7538 IRS\,1--3 region (in $JHK$ bands), 
and IRS\,9 region (in $HK$ bands) using the 8.2\,m Subaru telescope. The 
NIR analysis is complemented with GMRT low-frequency 
observations at 325, 610, and 1280\,MHz, molecular line observations of 
H$^{13}$CO$^{+}$ ($J$=1--0), and archival \textit{Chandra} X-ray observations. 
Using the \textquoteleft $J-H/H-K$\textquoteright\, diagram, 144 Class\,II 
and 24 Class\,I young stellar object (YSO) candidates are identified in the IRS\,1--3 
region. Further analysis using \textquoteleft $K/H-K$\textquoteright\, 
diagram yields 145 and 96 red sources in the IRS\,1-3 and IRS\,9 regions, respectively. 
A total of 27 sources are found to have X-ray counterparts. The YSO mass function 
(MF), constructed using a theoretical mass-luminosity relation, shows peaks at 
substellar ($\sim$0.08--0.18\msun\,) and intermediate ($\sim$1--1.78\msun\,) 
mass ranges for the IRS\,1--3 region. The MF can be fitted by a power law in the 
low mass regime with a slope of $\Gamma \sim$ 0.54-0.75, which is much shallower than 
the Salpeter value of 1.35. An upper limit of 10.2 is obtained for the star to brown 
dwarf ratio in the IRS\,1--3 region. GMRT maps show a compact \HII\, 
region associated with the IRS\,1--3 sources, whose spectral index of 
$0.87 \pm 0.11$ suggests optical thickness. This compact region is resolved into 
three separate peaks in higher resolution 1280\,MHz map, and the 
\textquoteleft East\textquoteright\, sub-peak coincides with the IRS\,2 source. 
H$^{13}$CO$^{+}$ ($J$=1--0) emission reveals peaks in both IRS\,1--3 and IRS\,9 
regions, none of which are coincident with 
visible nebular emission, suggesting the presence of dense cloud nearby. 
The virial masses are approximately of the order of 1000\msun\, and 500\msun\, for 
the clumps in IRS\,1--3 and IRS\,9 regions, respectively. 

\end{abstract}


\section{Introduction}  
NGC\,7538, an optically visible \HII\, region \citep[also known as Sh2-158]{fich84}, 
is a part of the Cas OB2 complex and located at $l=111.54^{o},~b=+00.78^{o}$, 
at a distance of 2.65\,kpc \citep{moscadelli09}. Early infrared observations of 
this region by \citet{wynn74} and \citet{werner79} revealed the presence of infrared 
sources IRS\,1-11 associated with and in the neighbourhood of the optical nebula. 
This region has been studied in prolific detail using various techniques such as 
high-frequency radio observations, line emission observations, maser observations, 
outflows, and so on. However, most of these studies as well as many recent ones 
% \citep{zhu13, beuther12, wright12, naranjo12, barentine12} 
concentrate on the 
individual features of luminous IRS and candidate high-mass sources. Even then, 
the bulk deals with the sources around IRS\,1--3 stellar cluster region, and the 
IRS\,9 region analyses have been sparse. Recently, an NIR study of the overall 
NGC\,7538 region by \citet{ojha04a}, reddening and cluster related studies using 
NIR by \citet{balog04}, a spectroscopic study of the luminous sources by 
\citet{puga10}, and a multiwavelength study of clusters (using statistical techniques) 
with a focus on high-mass stars by \citet{chavarria14} have been carried out. Other 
large scale studies have been carried out at far-infrared \textit{Herschel} bands 
and at submillimetre wavelengths to identify cold clumps and filamentary structures 
\citep{sandell04, fallscheer13}.  

According to \citet{ojha04a} and \citet{mccaughrean91}, there appear to be three distinct 
sub-regions which can be separated based on the stellar population and morphology, namely 
the IRS\,1--3, IRS\,4--8, and IRS\,9 sub-regions. 
Although the previous works identified a rich cluster membership in a wide ­field, encompassing 
all the major star-­forming sites in this complex, none of the optical/infrared surveys were 
deep enough to reach the substellar regime.  
Hence the works available in the literature were not able to have a detailed study of these 
stellar cluster sub-­regions, which need to be analysed separately due to their veritable differences. 
Recent advanced instruments on 8--10\,m class telescopes now make it 
feasible to conduct imaging and spectroscopic studies of low-mass populations in 
distant high-mass star-forming regions, where the star-forming environment may be 
different from their low-mass counterparts. Moreover, in the case of stars in a 
distant cluster, high resolution imaging is also needed to recognize individual 
stars. 
Since most of the available studies of low-mass populations are mainly based on nearby 
star-forming regions and the sample of substellar sources is small to draw definitive 
conclusions, we considered it worthwhile to focus on the stellar clusters of the NGC\,7538 
region observed using the 8.2\,m Subaru telescope - the deepest and highest-resolution NIR 
data till date. 
The NIR data has been used to study the luminosity function and initial mass function (IMF) 
of the region. 
In young massive star-forming regions, a gamut of components are present which could affect 
further evolution, and thus we have complemented our deep NIR observations with : 
previously unexamined low-frequency bremsstrahlung to ascertain the ionizing gas 
characteristics, molecular line H$^{13}$CO$^{+}$($J$=1 - 0) observations for the dense gas 
morphology, and \textit{Chandra} X-ray observations 
% (analysed for the first time, albeit limited) 
for stellar population analysis.   

In this paper, therefore, the aim is to continue with our multiwavelength study of 
star-forming regions 
\citep{ojha04a,ojha11,samal07,samal10}, as well as our investigations to detect and characterize the 
young brown dwarfs (BDs) in distant massive star­-forming regions \citep[cf.][]{ojha04b,ojha09}. 
In Section \ref{section_Obs_DataReduction}, we present details of the observations 
and data reduction procedures. The YSO selection procedure is dealt with in 
Section \ref{section_YSOselection}. The spatial distribution, morphology, and physical 
characteristics of the regions are discussed in Section \ref{section_morphology}. 
In Section \ref{section_LFandMF}, we elaborate upon the luminosity and mass functions 
obtained for different clusters. Discussion and final conclusions are presented in 
Sections \ref{section_discussion} and \ref{section_conclusion}, respectively.

\section{Observations and data reduction} 
\label{section_Obs_DataReduction}

\subsection{Near Infrared Photometry}
\label{section_NIRreduction}

Deep NIR imaging observations of the NGC\,7538 IRS\,1-3 region (centered on 
$\alpha_{2000}=23^{h}13^{m}43^{s}$, $\delta_{2000}=$ +61$^o$28\arcmin\,22\arcsec) 
in $J$ ($\lambda=$1.25 $\mu$m), $H$ ($\lambda=$1.64 $\mu$m), and 
$K$ ($\lambda=$2.21 $\mu$m) bands, and the NGC\,7538 IRS\,9 region 
(centered on $\alpha_{2000}=23^{h}13^{m}58^{s}$, 
$\delta_{2000}=$ +61$^o$27\arcmin\,26\arcsec) in $H$ and $K$ bands were obtained 
on 2005 August 19, using the Cooled Infrared Spectrograph and Camera for OHS (CISCO) 
mounted at the Cassegrain focus of the 8.2\,m Subaru telescope. These observations were 
done in the service mode of the telescope. CISCO is equipped 
with a 1024$\times$1024 Rockwell HgCdTe HAWAII array. A plate scale of 
0.105\arcsec\, pixel$^{-1}$ at the f/12 focus of the telescope provides a field-of-view 
(FoV) of $\sim$ 1.8\arcmin\, $\times$ 1.8\arcmin\, \citep{motohara02}. Observations 
were carried out in a 3$\times$3 dithering pattern with $\sim$ 10\arcsec\, offsets 
- with a varying number of images (three upto ten) being obtained at each dithered 
position. Individual image exposure times were 40\,s, 20\,s, and 10\,s for the 
$J$, $H$, and $K$ bands, respectively, finally yielding a total integration time 
of 12, 12, and 13.5 minutes in these respective bands. All the observations were 
done under excellent photometric sky conditions. The average seeing size was 
measured to be 0.4\arcsec\, full-width-at-half-maxima (FWHM) in all three filters, 
and the air mass variation was between 1.34 and 1.44. Off-target 
(located $\sim$ 34\arcmin\, east of the target position) sky-frame observations, 
identical in area to the target FoV, were taken just after the target observations 
using a similar procedure. 

Data reduction was done using the Image Reduction and Analysis Facility ({\sc iraf}) 
software package. The sky flats were generated by median-combining individual dithered 
sky frames for respective filters. These median-combined sky-flats were applied 
for both flat-fielding and sky subtraction. Identification of the point sources 
and their photometry was performed using the {\sc daofind} and {\sc daophot} packages of {\sc iraf}. 
Because of source confusion and nebulosity within the region, photometry was performed 
using the point spread function (PSF) algorithm {\sc allstar} in the {\sc daophot} package 
\citep{stetson87}. An aperture radius of 4 pixels ($\sim$ 0.4\arcsec\,) was used 
for the final photometry, with appropriate aperture corrections applied for the 
respective bands. Since no standard star was observed during the observations, the 
photometric calibration was carried out using sources (about 10-12, cross-matched 
within 0.4\arcsec\,of the Subaru catalogue) from the Two Micron All Sky 
Survey (2MASS)\footnote{This publication makes use of data products from the Two 
Micron All Sky Survey, which is a joint project of the University of Massachusetts 
and the Infrared Processing and Analysis Center/California Institute of Technology, 
funded by the National Aeronautics and Space Administration and the National Science 
Foundation.}, selected on the basis of their \textquoteleft phqual\textquoteright\, 
and \textquoteleft ccflg\textquoteright\, flag values, as well as visual examination, to 
avoid artifacts and contaminations.
Finally, the photometric calibration rms was $\sim$ 0.1\,mag. 
The Subaru/CISCO system magnitudes are assumed to be in California 
Institute of Technology (CIT) system \citep{oasa06}, and hence the 2MASS magnitudes 
too were converted to CIT system \citep[using ][]{carpenter01} for calibration purpose. 
On comparison of our photometry with that of \citet{ojha04a}, we find that 
the average dispersion (in the entire $K$ magnitude range) was $\sim$\,0.05-0.10\,mag for 
the $JHK$ bands. Our 
higher spatial resolution permits better source separation and sky determination.
Sources which were bright and saturated in Subaru images, but had  
good quality (\textquoteleft phqual=A or B\textquoteright\, for $J$, 
$H$, and $K_s$ bands each) 2MASS photometry had their magnitudes taken from the 
2MASS catalog after conversion to CIT system. Similar photometric procedure was 
also carried out for the off-target sky region. However, the western edge of the 
sky region was found to be slightly contaminated by a nearby nebula, and hence the 
sky frame was trimmed to remove the nebulous western portion before doing the 
photometry. The final sky frame size on which photometry (and completeness calculation 
below) was carried out is $\sim$ 1.17\arcmin\,$\times$1.8\arcmin\,. 
Since these observations were carried out in service mode, it was found that during 
the later observations, of IRS\,9 region in $K$ band, source profiles at the north-east 
part of the images were elongated, due to some likely optics problem. Since this part 
(north-east of 
IRS\,9 region) of the image is not crowded or nebulous, aperture photometry was carried 
out to estimate the source magnitudes of the sources with elongated profiles. Even then, 
due to limitations, we use IRS\,9 photomtery for qualitative assessments only in this work. 

\begin{figure*}
% \centering 
\includegraphics[trim={1.2cm 9.0cm 0cm 10.5cm}, clip, scale=1]{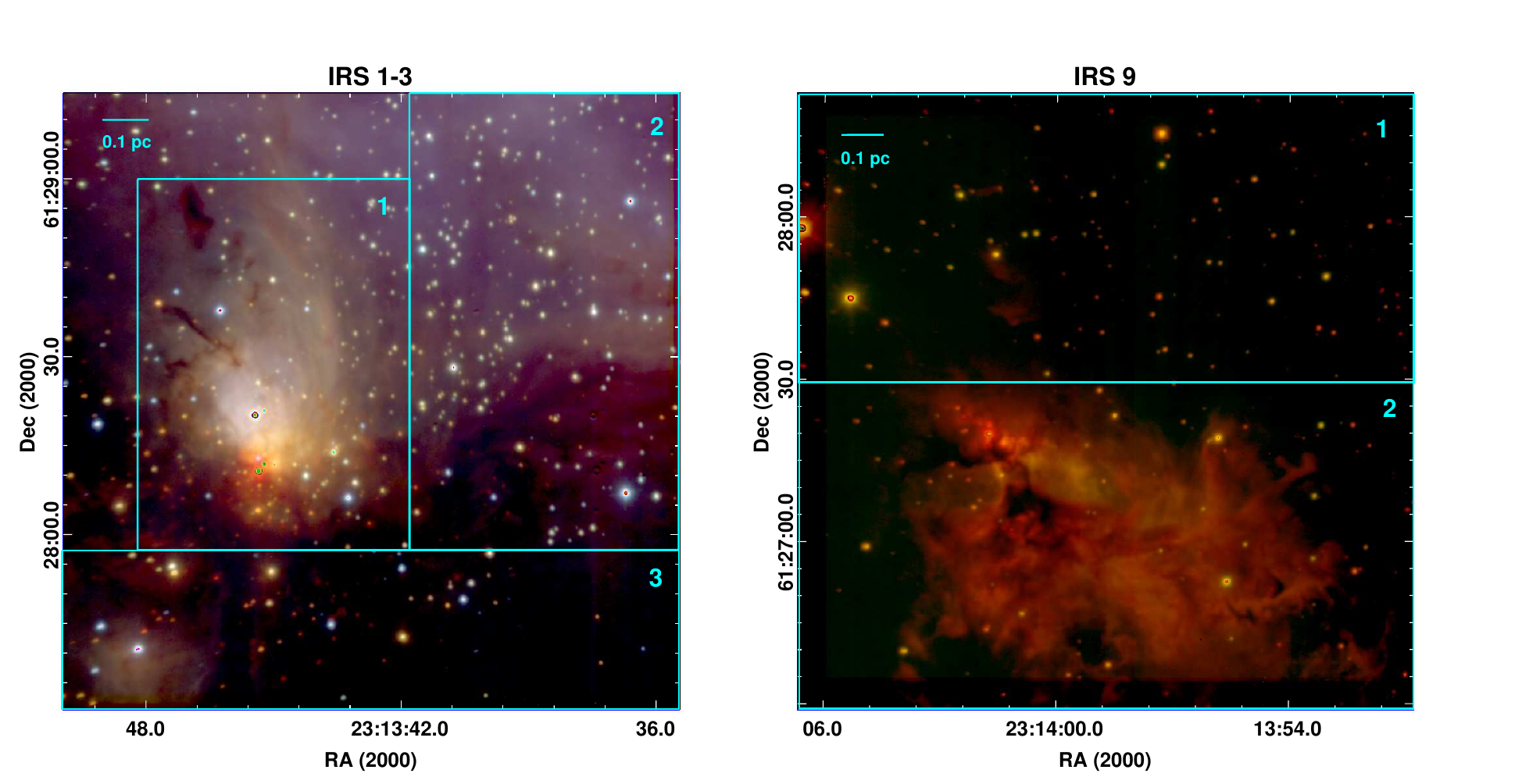}
\caption{\textit{(left)} Colour composite image of the IRS\,1-3 region using $J$ (blue), $H$ (green), and $K$ (red) 
bands. Three sub-regions are marked on the image using cyan rectangles. \textit{(right)} Colour composite image of 
the IRS\,9 region using $H$ (green), and $K$ (red) bands, with two sub-regions marked. Each sub-region of the respective 
region has been labelled by a number. Completeness limit was calculated for each sub-region.}   
\label{fig_ColourComposite_ForCompleteness} 
\end{figure*}

The completeness limits of the images were evaluated through artificial star 
experiments using {\sc addstar} in {\sc iraf}. Since the IRS\,1-3 and IRS\,9 regions have varying 
nebulosity, the images were divided into separate regions as shown in 
Figure \ref{fig_ColourComposite_ForCompleteness}, followed by completeness determination for 
each sub-region. A fixed number of stars were added in every 0.5 magnitude interval, 
followed by photometry to see how many of these added stars were recovered. This 
cycle was carried out repeatedly. We thus obtained the detection rate - which is 
just the ratio of the number of recovered artificial stars to the number of added 
stars - as a function of magnitude for each of the sub-regions in IRS\,1-3 and 
IRS\,9 regions, as well as the sky region. Table \ref{table_completeness} summarizes 
the 90\% completeness limits for all three bands in each of the sub-regions. The 
10$\sigma$ limiting magnitudes for our observations are estimated to be $\sim$ 22, 
21 and 20 in the $J$, $H$, and $K$ bands, respectively. As can be seen, the sky 
frame 90\% completeness limits are equal to these 10\,$\sigma$ limiting magnitudes 
for the $H$ and $K$ bands, and slightly lower for the $J$ band.

\subsection{Radio Continuum Observations} 
\label{section_RadioObs} 

Radio continuum observations were carried out using the Giant Metrewave Radio 
Telescope (GMRT) for the frequency bands 325 MHz (2004 July 03), 610 MHz (2004 
September 18), and 1280 MHz (2004 January 25). The GMRT array, consisting of 30 
antennae, is in an approximate Y-shaped configuration. Each of these antennae has 
a diameter of 45\,m, and thus a primary beam size of $\sim$ 81\arcmin\,, 
43\arcmin\,, and 26.2\arcmin\, for 325, 610, and 1280\,MHz, respectively\footnote{
GMRT manual from \\
http://gmrt.ncra.tifr.res.in/gmrt\_hpage/Users/doc/manual/ \\
Manual\_2013/manual\_20Sep2013.pdf}.  
There is a central region ($\sim$ 1\,km$\times$1\,km) which 
consists of randomly distributed 12 antennae, while the rest of the antennae are 
along three radial arms (6 along each arm) extending upto $\sim$ 14\,km each. The 
minimum and maximum baselines are 100\,m and 25\,km, respectively. Further details 
about the GMRT array can be looked up in \citet{swarup91}. 

The total observation time ranged from $\sim$ 2.5-3.5\,hours for the three frequency bands.
Data reduction was carried out using the {\sc aips} software. Successive rounds of flagging 
and calibration were carried out to improve the calibration. 
Flagging involved removal of bad 
data (including bad antennae, baselines, channels, time-ranges resulting from 
terrestrial radio frequency interference, etc), and was done using the 
\textquoteleft {\sc vplot-uvflg}\textquoteright\, and 
\textquoteleft {\sc tvflg}\textquoteright\, tasks. 
The respective flux and phase calibrators were used in the standard 
\textquoteleft {\sc calib-getjy-clcal}\textquoteright\, procedure 
for amplitude and phase calibration. After satisfactory calibration, 
the source data (NGC\,7538) was \textquoteleft {\sc split}\textquoteright\, from 
the whole file (which contains flux and phase calibrator data in addition). Facet 
imaging was done using the {\sc aips} task \textquoteleft {\sc imagr}\textquoteright\, 
to generate the requisite images. To remove the ionospheric phase distortion effects, 
a few rounds of (phase) self-calibration - with decreasing 
\textquoteleft solint\textquoteright\, - were carried out using the task 
\textquoteleft {\sc calib}\textquoteright\,. Table \ref{table_RadioObservation} 
gives the details of the GMRT observations and the parameters of the generated 
images. In addition, VLA archival image for 4860\,MHz, for the observation date 
2000 September 22 (Project ID BP0068), was also used in the analysis\footnote{This 
paper uses data produced as a part of the NRAO VLA Archive Survey (NVAS). The NVAS 
can be accessed through http://archive.nrao.edu/nvas/.}.

\subsection{\textit{Chandra} X-ray Observations}
\label{section_ChandraReduction} 

Publicly available archival X-ray data for NGC\,7538 region was retrieved from the 
\textit{Chandra} site\footnote{http://cda.harvard.edu/chaser/} (Obs. ID 5373). 
The X-ray observations of this \HII\, region were carried out using the Advanced 
CCD Imaging Spectrometer \citep[ACIS;][]{garmire03} onboard the 
\textit{Chandra X-ray Observatory} \citep[CXO;][]{weisskopf02}. For our purpose, 
we use only the imaging array of ACIS (ACIS-I). ACIS-I consists of a 2$\times$2 
CCD array of 1024$\times$1024 pixels each, with a total FoV of $\sim$ 
17\arcmin $\times$17\arcmin\,. The net exposure time was 30\,ks. 

Initial steps in the data reduction were carried out using the \textit{Chandra} 
Interactive Analysis of Observations \citep[{\sc ciao};][]{fruscione06} tool version 4.5 
and \textit{Chandra} Calibration Database ({\sc caldb}) version 4.5.8. The data was 
reprocessed using the \textquoteleft \textit{chandra\_repro}\textquoteright\, 
tool to apply the latest calibration to it. 
Light curve of source-free background regions were constructed to verify that 
the data was not affected by solar flare activity. 
Subsequently, with the help of the wavelet based source detection tool - 
\textquoteleft \textit{wavdetect}\textquoteright\, \citep{freeman02} - source 
detection was carried out at a threshold level of 2$\times$10$^{-6}$. Data image 
and exposure map with the default pixel scale of $\sim$ 0.5\arcsec\, were used for 
this purpose. Once we have a final list of positions of detected sources 
from \textquoteleft \textit{wavdetect}\textquoteright\,, we used 
the IDL based {\em ACIS Extract} 
({\sc ae}) software package\footnote{The {\em ACIS Extract} software package and User's 
Guide are available at http://www.astro.psu.edu/xray/acis/acis\_analysis.html.} 
\citep{broos10, broos12} version \textquoteleft March 6, 2013\textquoteright\, to 
extract the relevant source properties. The algorithm detailed in the {\sc ae} user's 
guide was followed. The source counts are extracted within $\sim$ 0.90 PSF fraction, 
with the PSF being separate for each source. 
Within {\sc ae}, the X-ray spectra are compiled and fitted with an optically-thin 
thermal plasma model attenuated by an interstellar absorption using the 
\textquoteleft {\sc xspec}\textquoteright\, fitting package. 
A total of 182 sources were obtained in the FoV, all of which were verified to 
have P$_B \leq$ 0.01, to make sure that the detected sources 
are real and not a result of Poissonian fluctuation in the local background. 
P$_B$ gives 
the probability that the extracted counts in the total band are solely a result 
of background fluctuations. 
The intrinsic hard band luminosity for all the sources was found to range from 
$\sim$ 5.5$\times$10$^{28}$ to 3.5$\times$10$^{32}$ erg\,s$^{-1}$, while the intrinsic total 
luminosity was in the range $\sim$ 5.5$\times$10$^{29}$ to 6$\times$10$^{35}$ erg\,s$^{-1}$.
The hard band luminosity function peak was at about 5$\times$10$^{30}$ erg\,s$^{-1}$.
For sources with spectral fitting results, the column density peaks at $\sim$\,10$^{22}$\,cm$^{-2}$
and the plasma temperature at $\sim$\,1\,keV. The brightest ones show some metallic 
emission lines, supporting the thermal origin of the X-rays. These typical features are
commonly seen among YSOs \citep{getman06}. 
We additionally used the non-parametric {\sc xphot} program of \citet{getman10} to estimate the 
intrinsic fluxes and X-ray column densities. 

Among the total number of sources obtained, there will also be extragalactic contaminants like 
Active Galactic Nuclei (AGN), and foreground stellar sources. Though we use a 
small subset of sources for our study (see Sections \ref{section_CCD} and \ref{section_CMD}), 
and thus have not carried out a detailed contaminantion analysis here, we can obtain an 
estimate using the values for the Cepheus\,B region \citep{getman06} which is close to NGC\,7538 
and had the same exposure time as well as FoV. 
\citet{getman06} find that the extragalactic contamination in Cepheus\,B is $\leq$\,5\%, 
while stellar contamination is $\leq$\,4\%. Assuming these numbers would imply a total 
of $\sim$\,16 contaminants (out of 182 total) in the NGC\,7538 ACIS-I FoV. 
Table \ref{table_XraySources} gives the properties of 27 X-ray sources with NIR counterparts 
in the NIR FoV which have been selected in this paper for further analysis.

\subsection{H$^{13}$CO$^{+}$($J$=1 - 0) Observations}
\label{section_H13CO+} 

The H$^{13}$CO$^{+}$ ($J$=1--0) (formylium) molecular line (86.754 GHz) 
observations were carried out on 2004 May 02 with the Nobeyama 45\,m radio 
telescope. At 87\,GHz, the half-power beam width and main beam 
efficiency, $\eta$, of the telescope were 18\arcsec~and 0.51, 
respectively. We used the 25-BEam Array Receiver System (BEARS) 
\citep{sunada00}. To correct for the beam-to-beam gain variation, we 
calibrated the intensity scale of each beam using a 100\,GHz SIS 
receiver (S100) with an SSB filter. Furthermore, we observed the same 
grid point in mapping with 9 different beams to smooth the beam-to-beam 
gain variation. At the back end, we used 25 arrays of 1024 channel 
Auto-Correlators (ACs), which have a 32 MHz band width and a 37.8 kHz 
resolution, corresponding to 0.13 km s$^{-1}$ \citep{sorai00}. All 
the observations were carried out in position-switching mode. The 
standard chopper wheel method was used to convert the received 
intensity into the antenna temperature, $T_{\rm{A}}^{*}$. Our 
mapping observations covered the same region of the NIR image. 
The mapping grid has 21\arcsec\, spacing, corresponding to half of 
the beam separation of the BEARS, i.e., nearly full-beam sampling. 
During the observations, the system noise temperatures were in the range 
of 200 to 400\,K, resulting in a noise level of 0.35\,K in $T_{{\rm A}}^{*}$.
The telescope pointing was checked every 1.5 hours by observing the 
SiO maser source R\,Cas. The pointing accuracy was better than 3$''$.

\section{YSO Selection} 
\label{section_YSOselection} 

\subsection{Using NIR Colour-Colour Diagram} 
\label{section_CCD} 

\begin{figure*}
\centering
\subfigure
{
\includegraphics[trim={1.5cm 7.0cm 0cm 8.5cm}, clip, scale=0.65]{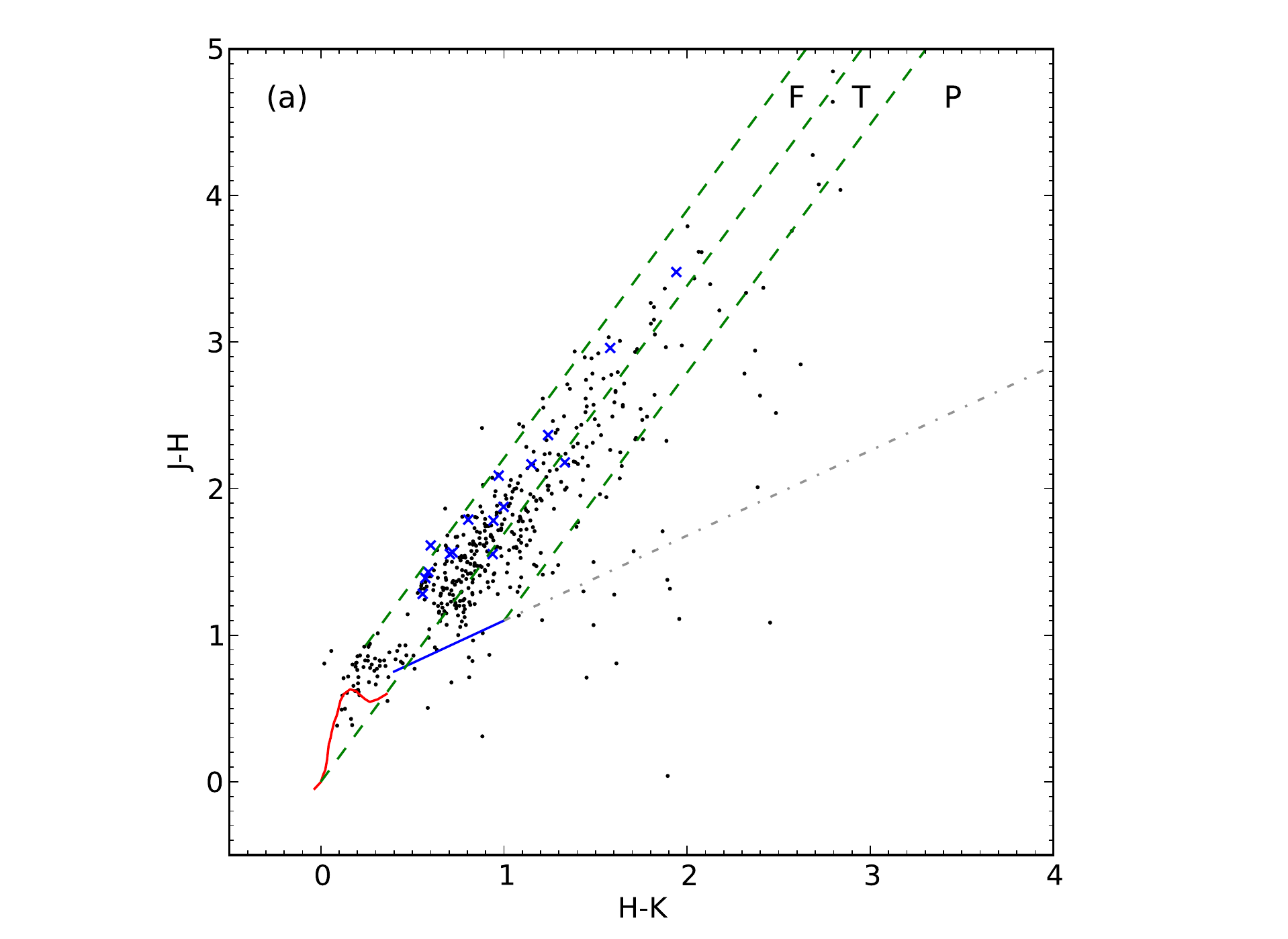}
\label{fig_NIRCCD_IRS13}
}
\subfigure
{
\includegraphics[trim={1.5cm 7.0cm 0cm 8.5cm}, clip, scale=0.65]{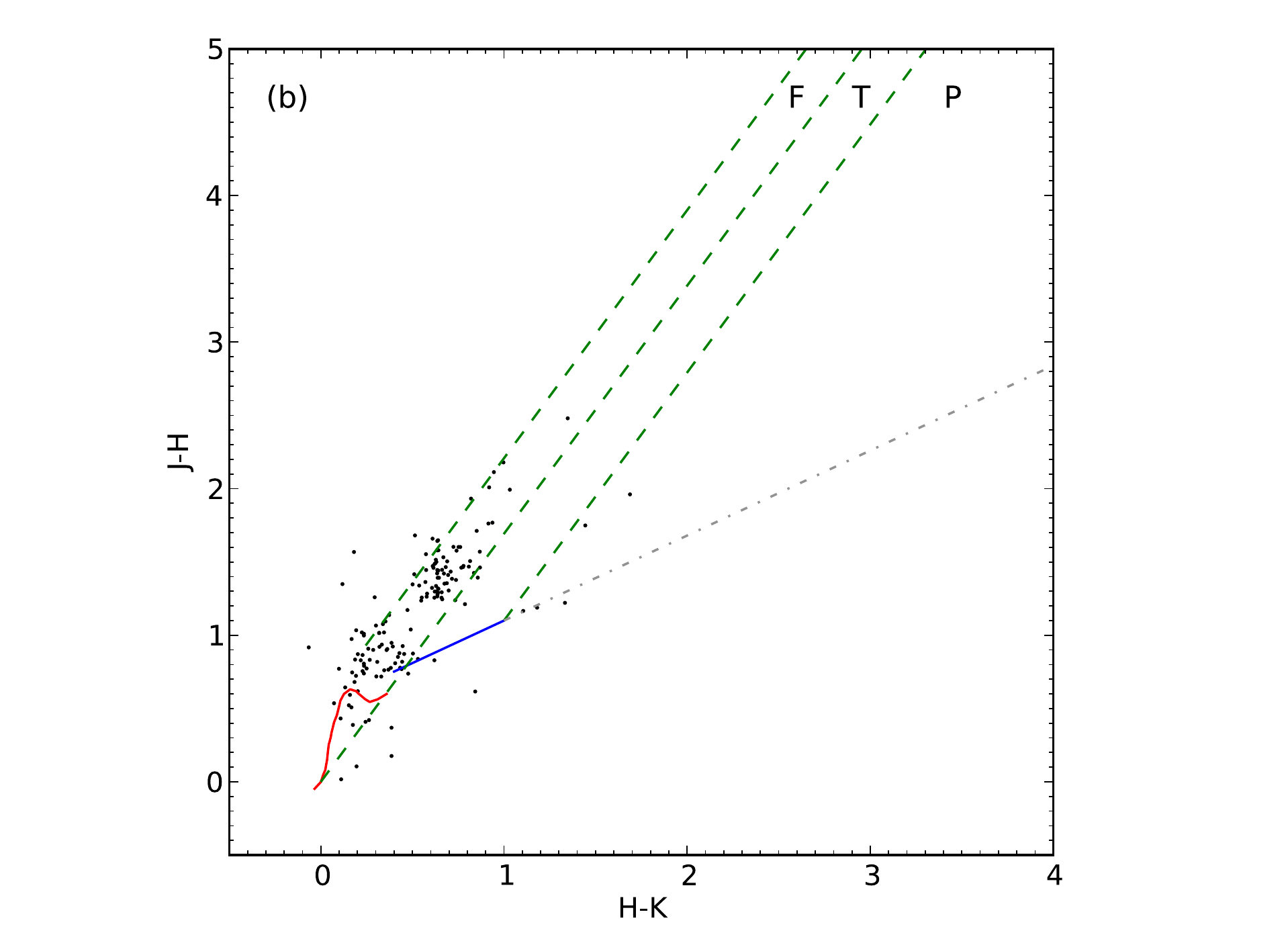}
\label{fig_NIRCCD_sky} 
}
\caption{(a) $J-H/H-K$ NIR CC-D for the IRS\,1-3 region. The red curve shows the dwarf locus from \citet{bessell88}. 
The CTTS locus \citep{meyer97} is shown by the blue solid line, and its extension into the 
\textquoteleft P\textquoteright\, region by grey dot-dashed line. The three parallel and slanted dashed lines are 
the reddening vectors using the reddening laws of \citet{cohen81}. Blue crosses mark the sources with X-ray counterparts. 
All points are in CIT system. (b) The CC-D for the sky field.}  
\label{fig_NIRCCD} 
\end{figure*}

The photometric catalogs containing the $J$, $H$, and $K$ band magnitudes for the IRS\,1-3 region (see 
Section \ref{section_NIRreduction}) and the sky field region were used to construct the NIR \textit{J-H vs H-K} 
colour-colour diagrams (CC-Ds) shown in Figure \ref{fig_NIRCCD}. In this figure, the red curve denotes the 
dwarf locus from \citet{bessell88} (converted to CIT system),  
blue solid line denotes the locus of Classical T Tauri Stars (CTTS) from \citet{meyer97}, and the three parallel 
dashed green lines denote the reddening vectors drawn using the reddening laws of \citet{cohen81} 
($A_J/A_V = 0.265, A_H/A_V = 0.155,$ and $A_K/A_V = 0.090$) for the CIT system. 
Three separate regions are marked on the images - \textquoteleft F\textquoteright\,, 
\textquoteleft T\textquoteright\,, and \textquoteleft P\textquoteright\, - similar to \citet{ojha04a,ojha04b}. 
The \textquoteleft F\textquoteright\, region mostly contains the field stars and Class\,III-type  
(Weak-line T Tauri or WTTS) sources, the  
\textquoteleft T\textquoteright\, region mostly contains the CTTS and Class\,II-type sources \citep{lada92}, and 
the \textquoteleft P\textquoteright\, region mostly contains the Class\,I-type sources with circumstellar envelopes. 
Since the sources in the \textquoteleft P\textquoteright\, region can be contaminated by 
Herbig Ae-Be stars \citep{lada92}, we conservatively only consider those sources in this region whose \textit{J-H}
colour is larger than that of the CTTS locus extended into this region. There may also be an overlap in the NIR 
colours of the upper end band of Herbig Ae-Be stars and in the lower end band of CTTS in the \textquoteleft T\textquoteright\, 
region \citep{hillenbrand92}. 

The sky field CC-D shown in Figure \ref{fig_NIRCCD_sky} is used to examine the extent upto which IRS\,1-3 CC-D is 
affected by field star contamination. As can clearly be seen in Figure \ref{fig_NIRCCD_sky}, almost all field star 
contamination is present in the 
\textquoteleft F\textquoteright\, region. Therefore, while the sources falling in the \textquoteleft T\textquoteright\, and 
\textquoteleft P\textquoteright\, regions of Figure \ref{fig_NIRCCD_IRS13} are most probably YSOs, the Class\,III-type 
sources from the \textquoteleft F\textquoteright\, region will be contaminated by field stars. 

To further separate the most-likely YSOs from the \textquoteleft F\textquoteright\, region, 
since pre-main sequence (PMS) stars display much stronger X-ray emission than the field main sequence (MS) 
stars \citep{feigelson99,montmerle96}, we use the X-ray source identifications from 
Section \ref{section_ChandraReduction}. 
For this, we cross-matched our NIR catalog with the X-ray catalog within 0.5\arcsec\,radius. 
X-ray detected sources, however, in general, suffer from extragalactic 
contamination (mostly AGN) and foreground stellar contamination. 
But since we are using only those X-ray sources which have NIR counterparts, it is unlikely - similar to 
the work of \citet[for Cepheus\,B]{getman06} and \citet[for W40]{kuhn10} - that there will be any extragalactic 
contamination. Also since the resolution of \textit{Chandra} images ($\sim$ 0.5\arcsec\,) is similar to our NIR imaging 
($\sim$ 0.4\arcsec\,, see Section \ref{section_NIRreduction}), mismatching should not be a concern. 
From the matching radius and the source number density, the number of X-ray sources to have an NIR counterpart by chance 
was calculated to be one at most. 
Apropos foreground contamination, the X-ray sources lie beyond the low-density gap at about $H-K \sim 0.5$ 
\citep[again similar to][]{kuhn10} in the NIR diagrams. Normally, a sudden low-density gap 
signifies the boundary of foreground sources in $H-K$ colour space. Hence, these X-ray matched sources are 
unlikely to be foreground contaminants too.

Finally - using the CC-D for IRS\,1-3 region - 251 Class\,III-type sources including probable contaminants (14 have X-ray 
counterparts), 144 Class\,II-type sources (2 have X-ray counterparts), and 24 Class\,I-type YSO candidates were 
identified from the \textquoteleft F\textquoteright\,, 
\textquoteleft T\textquoteright\,, and \textquoteleft P\textquoteright\, regions, respectively. 
Table \ref{table_YSOs_IRS13} lists these YSO candidates along with their 
respective NIR magnitudes, and IAU designation from Table \ref{table_XraySources} where applicable. 
Similar NIR CC-D could not be constructed for the IRS\,9 region as we only have $H$ and $K$ magnitudes available 
for it.

\subsection{Using NIR Colour-Magnitude Diagram} 
\label{section_CMD} 

\begin{figure*}
\centering
\subfigure
{
\includegraphics[trim={3.5cm 7.0cm 3.5cm 7.5cm}, clip, scale=0.57]{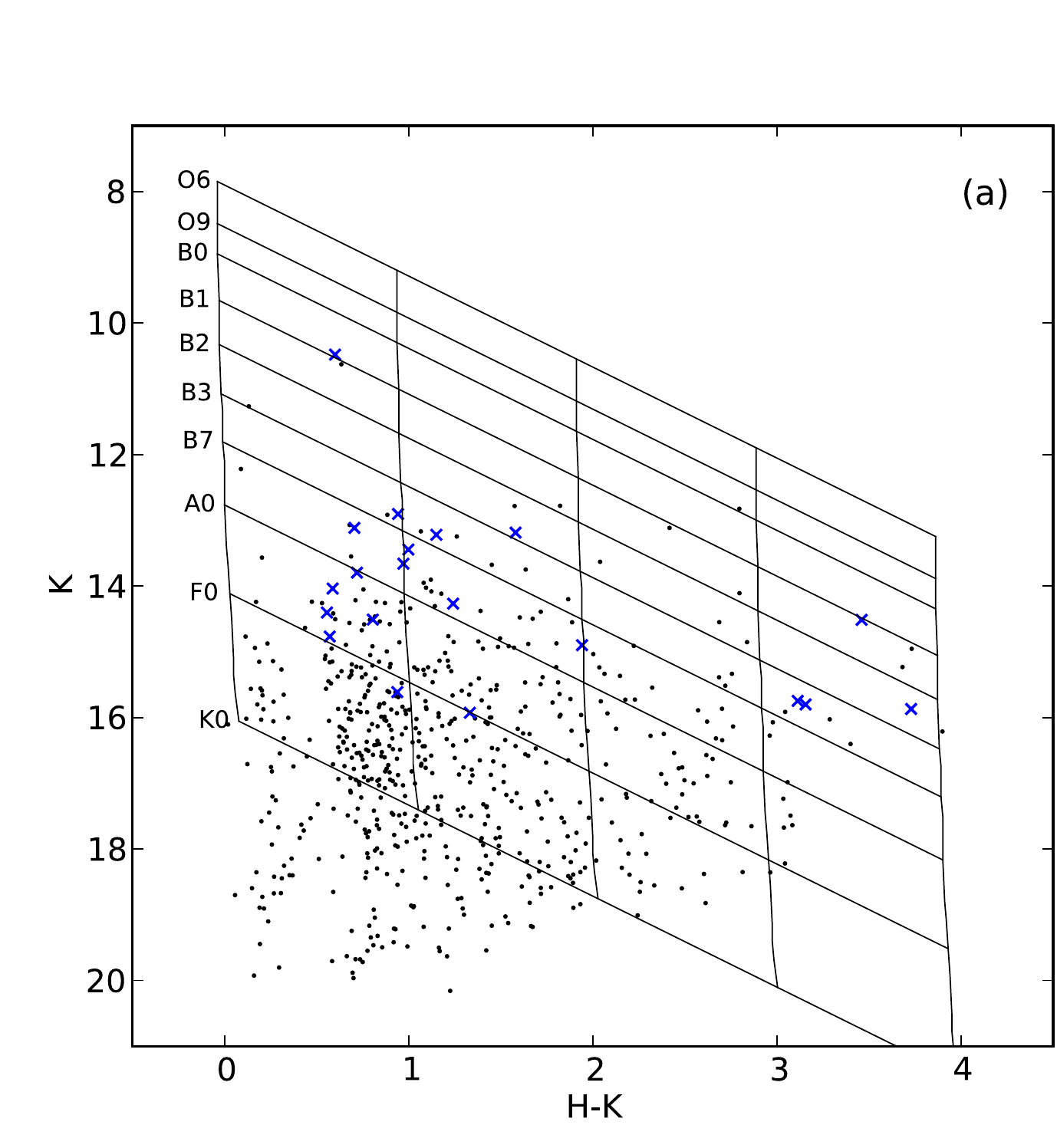}
\label{fig_NIRCMD_IRS13}
}
\subfigure
{
\includegraphics[trim={3.5cm 7.0cm 3.5cm 7.5cm}, clip, scale=0.57]{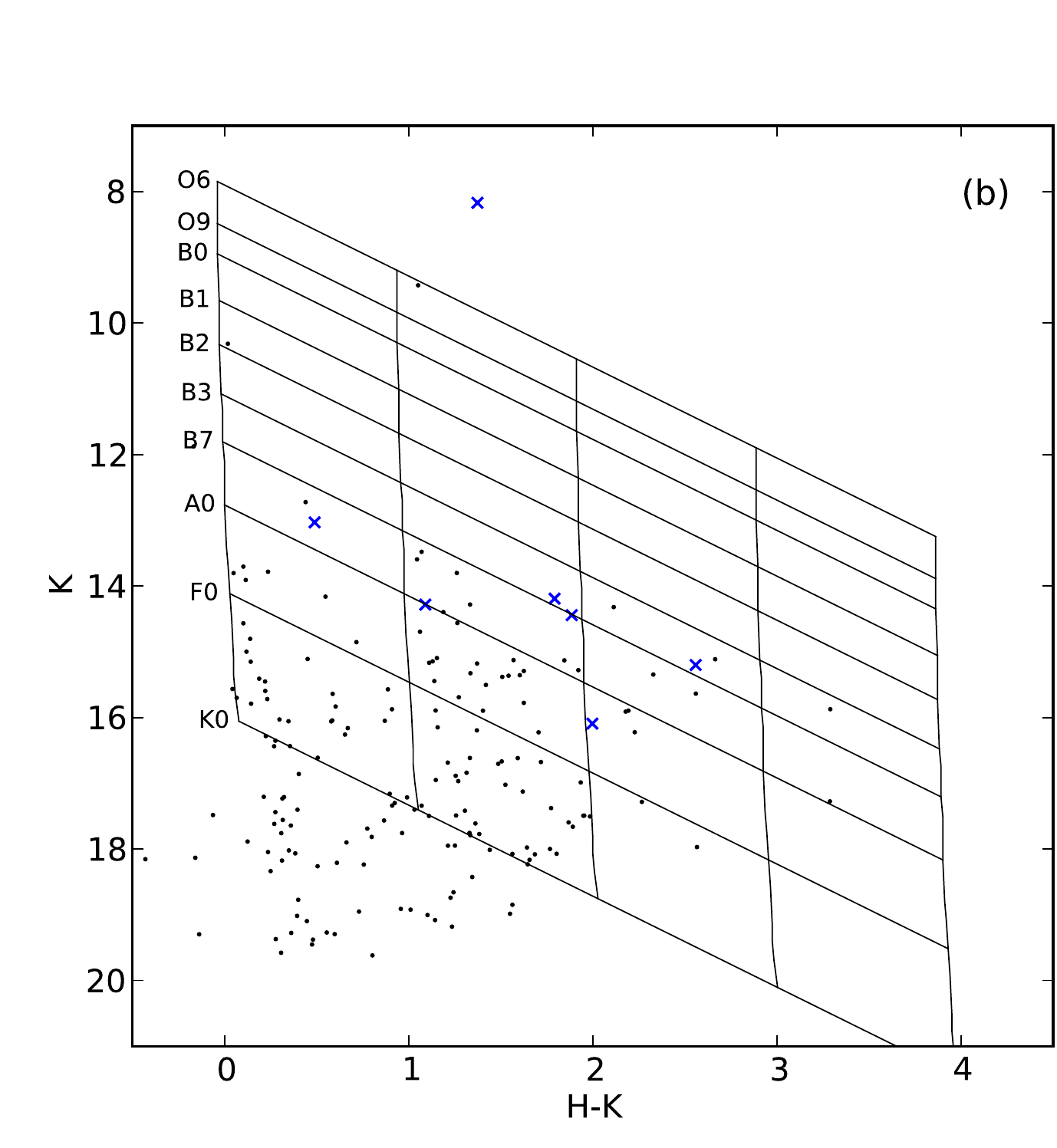}
\label{fig_NIRCMD_IRS9} 
}
\subfigure
{
\includegraphics[trim={3.5cm 7.0cm 3.5cm 7.5cm}, clip, scale=0.57]{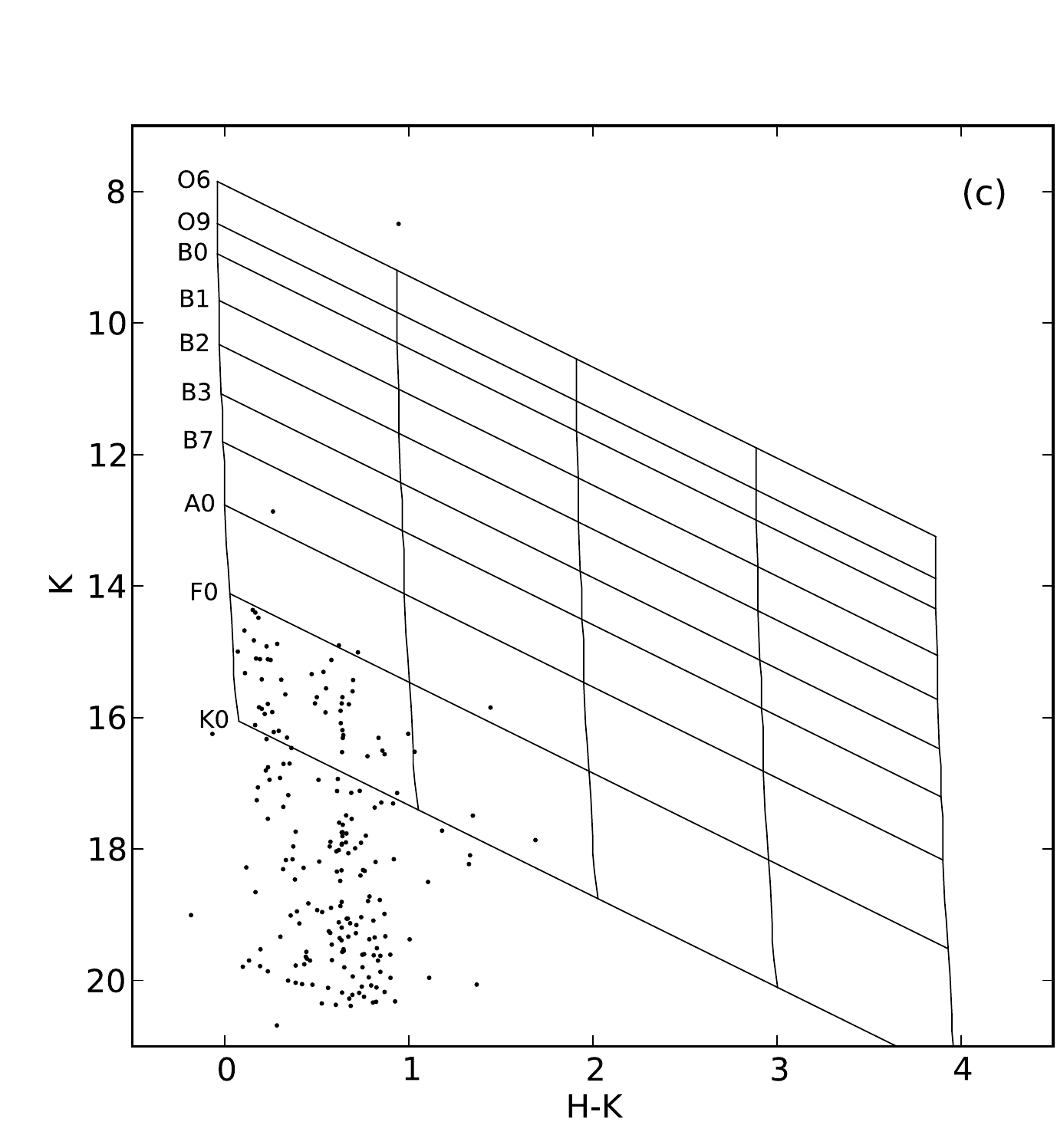}
\label{fig_NIRCMD_sky} 
}
\caption{$K/H-K$ NIR CM-D for (a) IRS\,1-3 region, (b) IRS\,9 region, and (c) the sky region. Nearly 
vertical solid lines are the loci of ZAMS stars reddened by $A_{V}$=0, 15, 30, 45, and 60 mag. 
Parallel, slanting lines indicate the reddening vectors for respective spectral types. Blue crosses 
mark the sources with X-ray counterparts.}   
\label{fig_NIRCMD} 
\end{figure*}

Many embedded and young sources can only be seen in $H$ and $K$ bands due to high extinction at $J$ band wavelengths. 
Therefore we use the \textit{K/H-K} 
colour-magnitude diagram (CM-D) for the identification of additional YSOs. Figure \ref{fig_NIRCMD} shows the \textit{K/H-K} 
CM-D for IRS\,1-3, IRS\,9, and the sky field regions. The almost vertical solid lines represent the zero-age-main-sequence (ZAMS)
locus at a distance of 2.65\,kpc reddened by $A_V = $ 0, 15, 30, 45, and 60 mag. Slanting lines indicate the reddening 
vectors for the marked spectral types. A low density gap in $H-K$ colour $\sim 0.5$ can be seen in all three diagrams, 
though it is much less pronounced for the IRS\,9 region due to the lack of statistics. 
In the sky field CM-D (Figure \ref{fig_NIRCMD_sky}), we can see that most of the sources are confined to $H-K \leq 1$. 
This $H-K$(=1) limit also corresponds to the average extinction of $A_V = 15$ mag found towards the NGC\,7538 region 
by \citet{ojha04a}. 
Thus, in general, a background contaminant field source suffering extinction due to the 
cloud should also be confined to $H-K \leq 1$. 
Also, on the same lines as \citet{ojha04a}, if we were to assume that sources had large IR colours (say, $H-K>2$) purely 
due to interstellar extinction, then that would imply that the reddening due to molecular cloud is $A_V >$\,30\,mag. 
But, with such a large $A_V$, diffuse emission will hardly be seen in NIR, which is definitely not the case 
here. 
Sources with colour over and above this limit (i.e. $H-K > 1$) are therefore most likely to have 
large $H-K$ colour due to intrinsic infrared (IR) excess associated with YSOs. 
Hence, we use this colour cut-off of $H-K > 1$ to identify extra YSO candidates from 
the IRS\,1-3 and IRS\,9 regions. To distinguish the YSO candidates identified using the NIR CM-D from those by the 
CC-D (Section \ref{section_CCD}), we refer to these sources as \textquoteleft red sources\textquoteright\, 
throughout the text. 145 sources (4 X-ray counterparts) were identified in the IRS\,1-3 region, and 95 (7 X-ray 
sources with $H$ and $K$ counterparts $+$ 1 X-ray source with only $K$ counterpart) in the IRS\,9 region. The catalogs of 
these red sources for the IRS\,1-3 and IRS\,9 regions are included 
in Tables \ref{table_YSOs_IRS13} and \ref{table_YSOs_IRS9}, respectively.

\section{Morphology and Spatial Distribution}  
\label{section_morphology}

\subsection{IRS\,1--3 region} 
\label{section_morphology_IRS13}

\begin{figure*}
\centering 
\includegraphics[trim={1.5cm 4.0cm 0cm 5.5cm}, clip, scale=0.7]{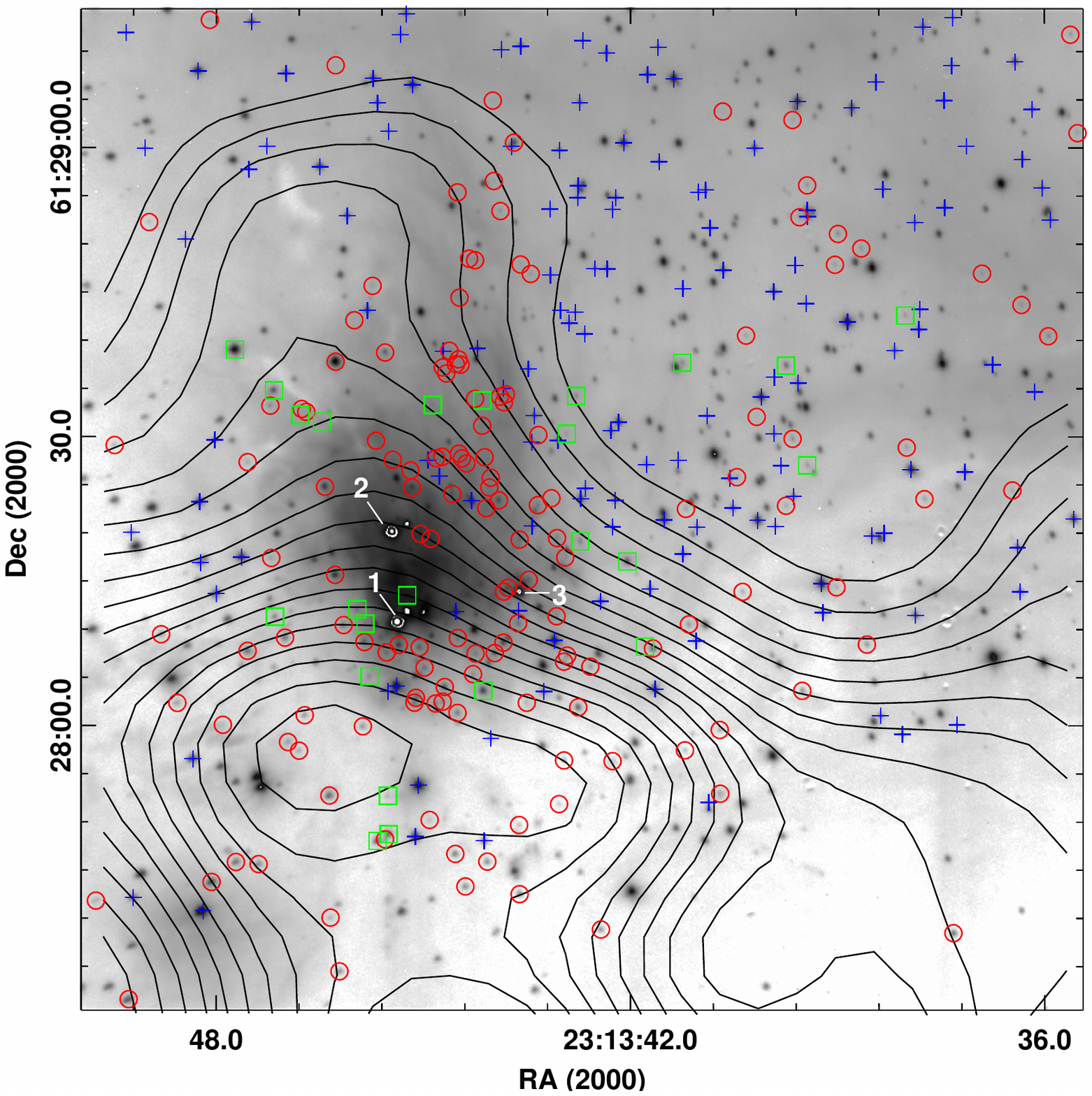}
\caption{$K$ band image of the IRS\,1-3 region overlaid with YSOs and H$^{13}$CO$^{+}$($J$=1--0) contours (black lines).
The contours are drawn in the range 0.5--2.2\,K\,km\,s$^{-1}$, with step size of 0.1\,K\,km\,s$^{-1}$. Class\,I sources 
are shown with green square symbols, Class\,II with blue plus symbols, and sources with $H-K>1$ with red circles. IRS\,1, 
2, and 3 are saturated in the image and have been marked.}   
\label{fig_Morphology_IRS13} 
\end{figure*}

The $K$ band image of the IRS\,1--3 region is shown in Figure \ref{fig_Morphology_IRS13} with the overlaid YSOs 
and H$^{13}$CO$^{+}$($J$=1--0) contours. Green squares mark the Class\,I sources, blue plus symbols the 
Class\,II-type or CTTS, and red circles the sources with $H-K>1$. IRS\,1, 2, and 3 are saturated and have been marked. 
Dense cloud is seen around these marked IRS sources. There appears to be relatively higher nebulosity towards the 
northern as opposed to the southern region. Almost all stellar sources are present in the northern region. While 
the Class\,II-type sources (blue plus symbols) seem distributed throughout the northern part, sources with 
$H-K>1$ (red circles) and Class\,I sources (green squares) are mostly concentrated around the dense cloud 
surrounding the luminous IRS sources. 
Some of them could be PMS members of the embedded stellar clusters associated with the molecular clumps around 
IRS\,1-3 and to the south of IRS\,1-3. A few red sources are also seen around the ionization front 
(the boundary between darker nebulous region and the lighter non-nebulous region in the western part; 
see Section \ref{section_RadioAnalysis} too) at the interface 
between the \HII\, region and the molecular cloud, which might have formed due to triggered star formation. 
The large diffuse emission extending to the north-west of IRS sources is probably due to a combination of free-free
and bound-free emissions, corresponding to what is seen optically, and coincides well with the radio brightness 
from the GMRT observations, while the bright and compact infrared nebula embedded within IRS\,1, 2, and 3 is 
coincident with the peak of radio continuum (see Section \ref{section_RadioAnalysis}).

\begin{figure*}
\centering
\subfigure
{
\includegraphics[trim={0cm 0.0cm 0cm 12.5cm}, clip, scale=0.6]{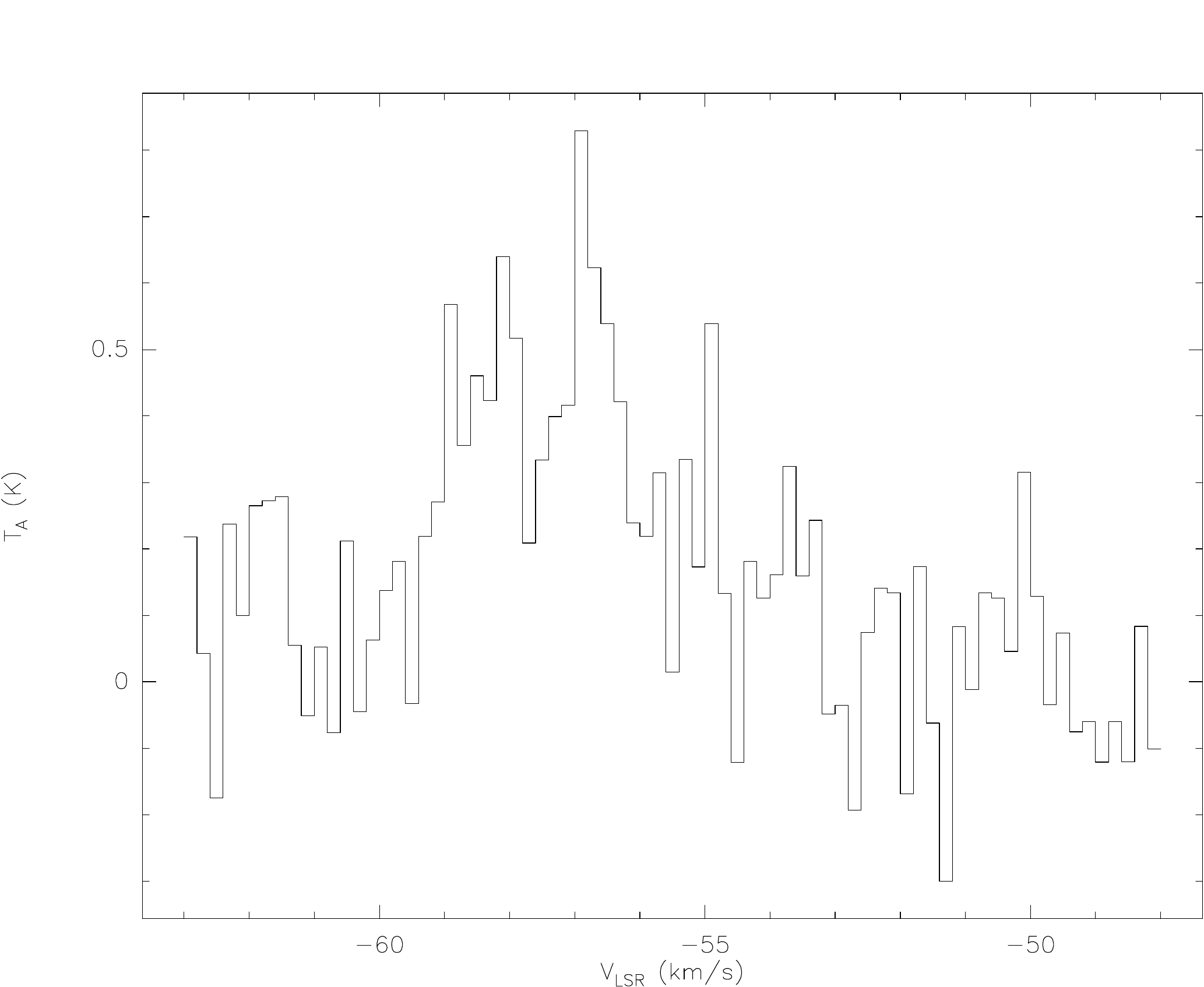}
\label{fig_H13CO+_IRS13}
}
\subfigure
{
\includegraphics[trim={0cm 0.0cm 0cm 12.5cm}, clip, scale=0.6]{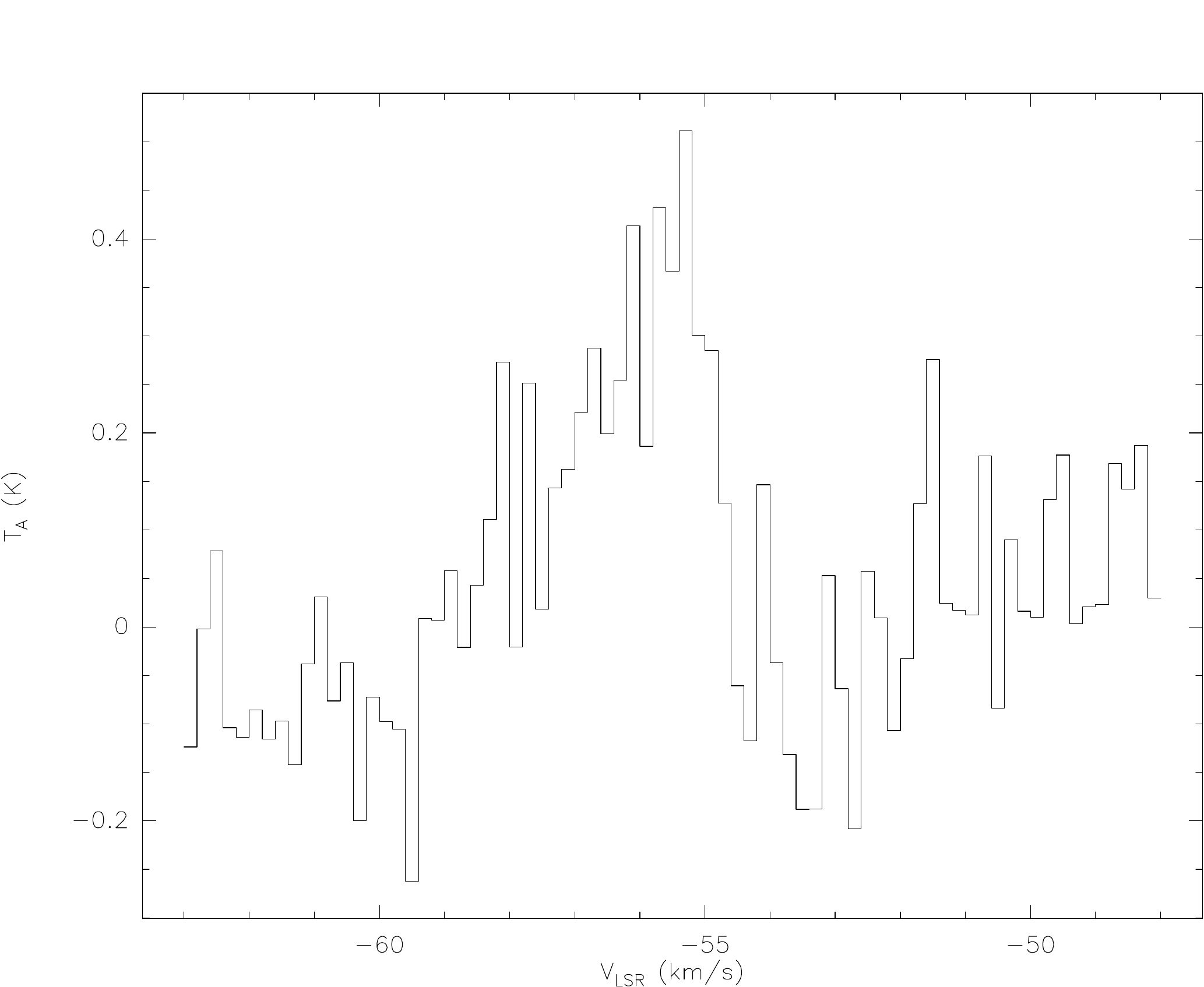}
\label{fig_H13CO+_IRS9} 
}
\caption{The H$^{13}$CO$^{+}$($J$=1--0) molecular line spectra towards \textit{(upper)} the peak in IRS\,1--3 region, and 
\textit{(lower)} the peak in IRS\,9 region.}    
\label{fig_H13CO+} 
\end{figure*}

The H$^{13}$CO$^{+}$($J$=1--0) contours show a peak to the south-east of the IRS\,1--3 nebula, where YSO density is 
much lower, and most of the sources around and near this peak are the red $H-K>1$ sources. Since this molecular 
emission traces the dense molecular cloud, it seems reasonable indeed that relatively very few sources are detected 
in the southern portion where the peak and the higher intensity contours lie, and the few sources detected are very 
reddened ones. The northern portion shows a hump in H$^{13}$CO$^{+}$($J$=1--0) contours, with contours closely 
spaced as one moves from the northern hump towards the southern peak. 
Preliminary calculations were carried out to get an idea of the column densities, local thermodynamic equilibrium (LTE)
mass, and virial mass of the clump associated with the peak. 
Assuming LTE, we tried to estimate the molecular column density using the formula from \citet{troitsky05}, 
with an excitation temperature of 10\,K, and a dipole moment of 3.9\,debye \citep{botschwina93}. 
Fitting a 2D Gaussian to the clump results in a peak integrated line intensity of 2.0\,K\,km\,s$^{-1}$ and a source size 
at half intensity level of 1\,pc. 
This gives us the LTE column density n(H$^{13}$CO$^{+}$) as $\sim 3.6 \times 10^{12}$\,cm$^{-2}$ after applying the 
main beam efficiency correction. Since the abundance - X(H$^{13}$CO$^{+}$) - 
estimates are variable from region to region and we do not have a measure of it here, we use the range seen for 
other massive star-forming regions - 0.5--3.0$\times$10$^{-10}$ from \citet{zinchenko09}. Using these values, n(H$_2$) range is 
obtained to be $\sim$1.2--7.2$\times$10$^{22}$\,cm$^{-2}$, and the LTE mass range of the clump to be $\sim$250--1500\msun\,. 
The spectrum for the IRS\,1--3 region clump is shown in Figure \ref{fig_H13CO+}\textit{(upper)}. 
Applying a 1D fit to the spectrum of the clump, we obtain a linewidth of $\sim$ 3.4\,km\,s$^{-1}$. The source size 
is corrected for the beam width to get the \textquoteleft deconvolved source size\textquoteright\, 
\citep{zinchenko95}. Using these values along with the expression from \citet{zinchenko94}, the virial mass is 
approximated to be of the order of 1000\msun\,. Alternatively, under the reasonable assumption of gravitational equilibrium, 
the LTE mass and virial mass can be equated, which will imply an abundance X(H$^{13}$CO$^{+}$) of 
$\sim$ 0.74$\times$10$^{-10}$ - which is within the range derived by \citet{zinchenko09}. 

\begin{figure*}
\centering
\includegraphics[trim={1.5cm 7.0cm 0cm 7.5cm}, clip, scale=0.5]{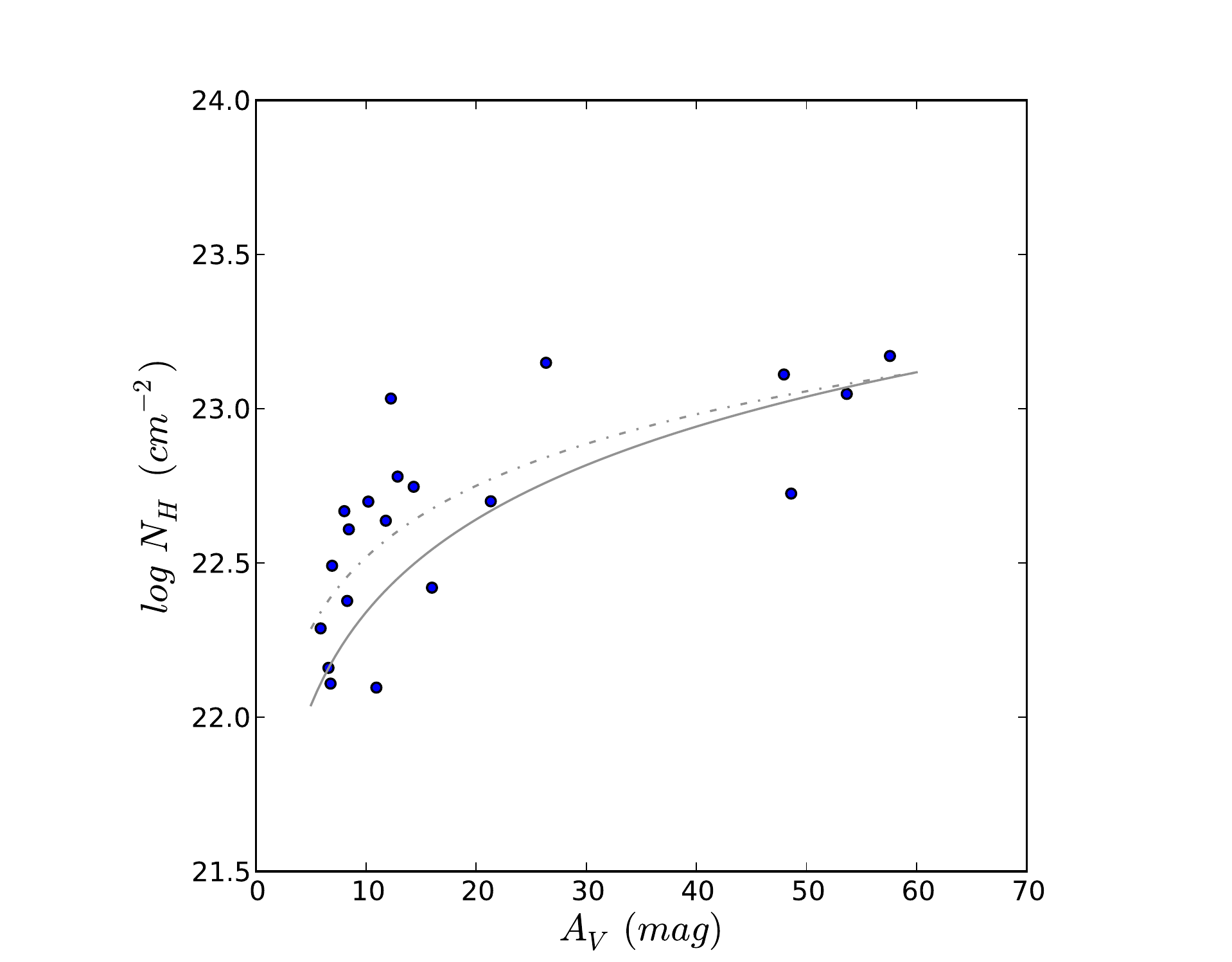}
\caption{X-ray column density ($\log N_H$) vs visual extinction ($A_V$) plot for the sources in the IRS\,1--3 region. 
The solid grey line marks the \citet{ryter96} gas-to-dust relation of $N_H = 2.2 \times 10^{21} A_V$, while the 
dash-dotted grey line marks the relation from our analysis.}   
\label{fig_logNH_Av} 
\end{figure*}

The gas-to-dust ratio for a cluster can be estimated by the relation obtained between the X-ray column density and 
the visual extinction for each source \citep{kuhn10}. Using the value of this column density ($log N_{H1}$) from 
Table \ref{table_XraySources} and $A_V$ from IR analysis (discussed in Section \ref{section_MF_IRS13}), we find 
$\log (N_H/A_V) \sim 21.75 \pm 0.20$. The relation is shown in Figure \ref{fig_logNH_Av}. It is slightly higher 
than that for the interstellar medium, $\sim$ 21.34 from \citet{ryter96}, and other young star-forming regions such 
as W40 \citep{kuhn10}. This higher value of gas-to-dust ratio is probably suggestive of dense gas in the 
region, as is also suggested by radio analysis (see Section \ref{section_RadioAnalysis}). 

\subsubsection{Stellar Cluster Analysis} 
\label{IRS13_stellarcluster} 

\begin{figure*}
\centering
\subfigure
{
\includegraphics[trim={1.5cm 4.0cm 2cm 5.5cm}, clip, scale=0.47]{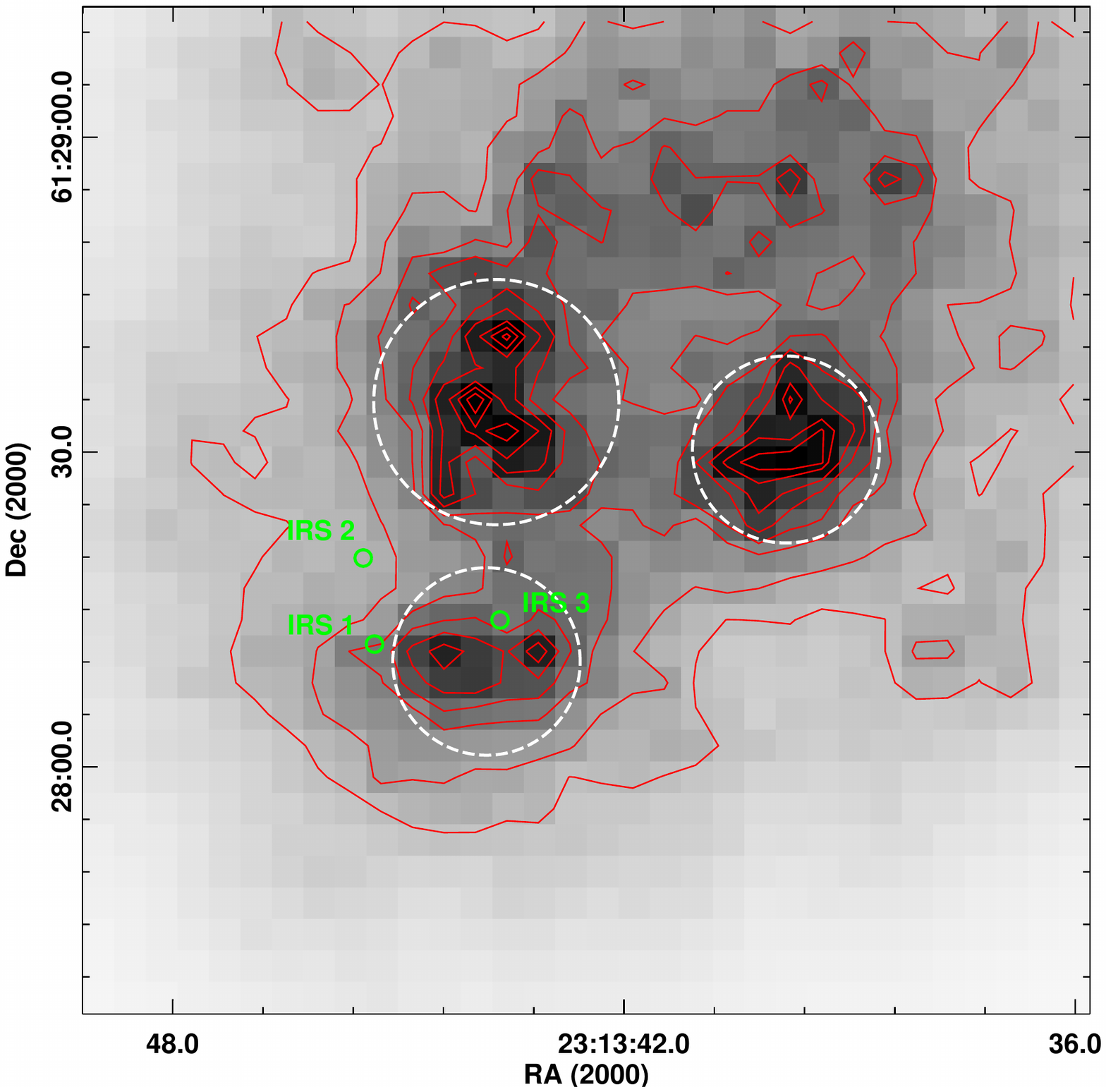}
\label{fig_IRS13_clusterings}
}
\subfigure
{
\includegraphics[trim={3.5cm 6.6cm 3cm 4.5cm}, clip, scale=0.6]{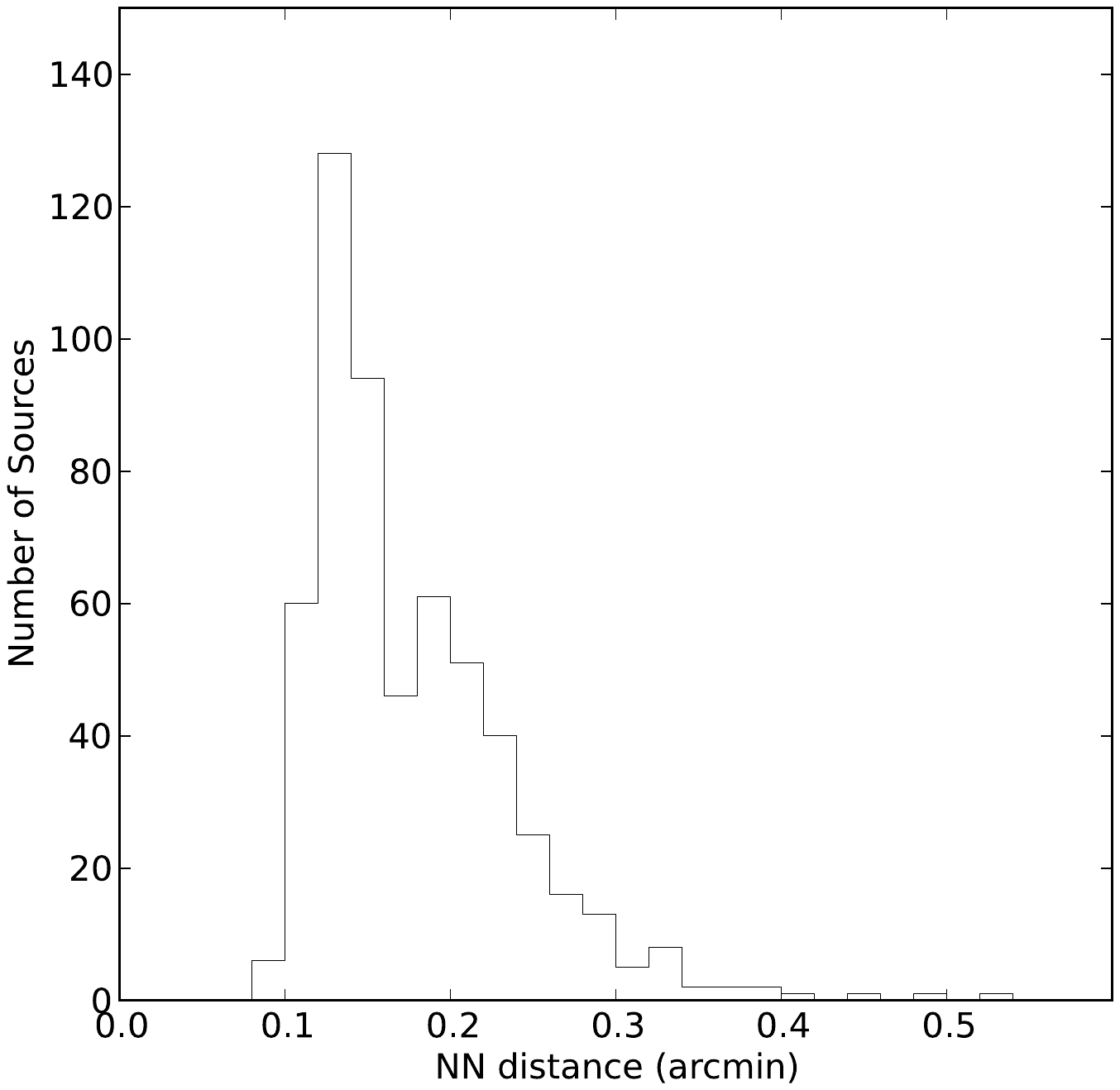}
\label{fig_IRS13_NN_histogram} 
}
\caption{\textit{(left)} 20\,NN surface density map of the IRS\,1-3 region with overplotted contours. 
The IRS sources have been marked by small green circles and labelled. 
The contour levels are at 250, 350, 500, 600, 700, 800, 900, 950, and 1000 YSOs pc$^{-2}$. Three distinct clusterings 
(indicated by white dashed circles) 
can be seen on the image - one close to the IRS\,1 and IRS\,3 sources, while the other two to the north and north-west of the 
IRS\,3 source. All display multiple peaks. \textit{(right)} The histogram of the NN distances of the sources.}  
\label{fig_IRS13_cluster} 
\end{figure*}

A stellar surface density analysis of this region was carried out using the nearest-neighbour (NN) method 
\citep{casertano85,schmeja08}. We use the catalog of candidate YSOs identified in Section \ref{section_YSOselection}, and 
a procedure similar to \citet{schmeja08} with 20 NN. The resultant surface density map is shown in 
Figure \ref{fig_IRS13_cluster}\textit{(left)}, also overplotted with contours. 
Figure \ref{fig_IRS13_cluster}\textit{(right)} shows the histogram of NN distances, with the peak NN distance in 
0.12-0.14\arcmin\, range. 
In Figure \ref{fig_IRS13_cluster}\textit{(left)} three major clusterings can 
be made out. One is towards the south, associated with the IRS\,1 and IRS\,3 sources, while the other two are to the 
north and north-west of the IRS\,3 source. Each of these clusterings display multiple peaks of their own. The maximum 
surface density values range from 860 to 1050 pc$^{-2}$, with the maximum exhibited by the clustering to the 
north of the IRS\,3. These values are higher than the peak calculated for the overall NGC\,7538 region by \citet{chavarria14},
which could be a result of deeper data used here. These high peak values, though, are similar to that for 
Serpens Core(A) \citep[1045 pc$^{-2}$; ][]{schmeja08}.

\subsection{IRS\,9 region} 
\label{section_morphology_IRS9}

\begin{figure*}
\centering 
\includegraphics[trim={1.5cm 4.0cm 0cm 5.5cm}, clip, scale=0.7]{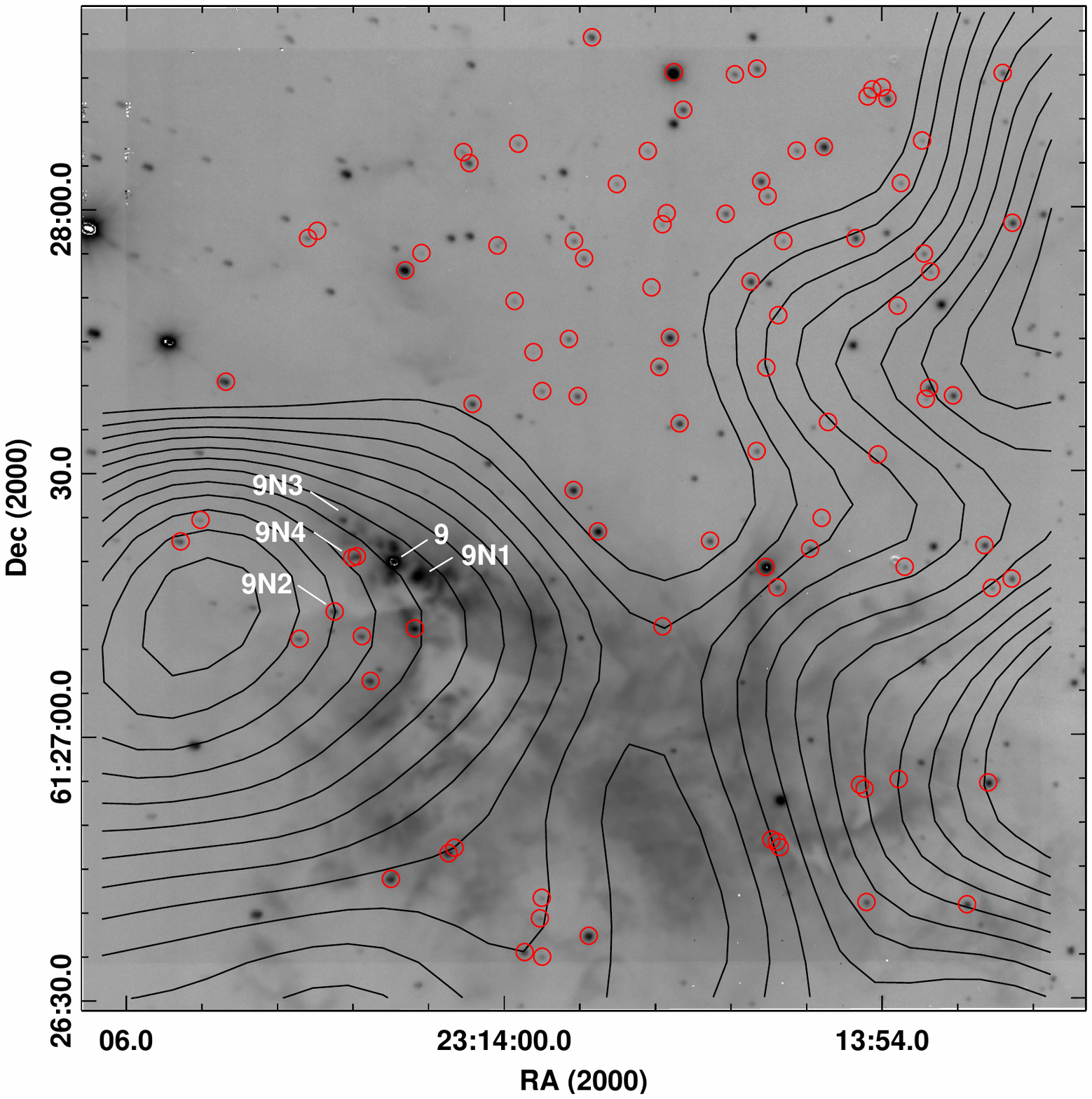}
\caption{$K$ band image of the IRS\,9 region overlaid with YSOs and H$^{13}$CO$^{+}$($J$=1--0) contours (black lines).
The contours are drawn in the range 0.5--1.6\,K\,km\,s$^{-1}$, with step size of 0.1\,K\,km\,s$^{-1}$. 
Candidate YSOs have been shown with red circles. 
IRS\,9, 9N1, 9N2, 9N3, and 9N4 sources from \citet{ojha04a} have been marked.}   
\label{fig_Morphology_IRS9} 
\end{figure*}

Figure \ref{fig_Morphology_IRS9} shows the $K$ band image of the IRS\,9 region with the overlaid YSOs 
and H$^{13}$CO$^{+}$($J$=1--0) contours. Red circles denote YSO candidate sources. 
The main IRS sources - 9, 9N1, 9N2, 9N3, and 9N4 - from \citet{ojha04a} - 
have been marked on the image. As can be seen in Figure \ref{fig_Morphology_IRS9}, sources are distributed throughout 
the image in no particular orientation, but with significantly more sources in the northern portion as opposed to the 
nebular southern portion - possibly an indication of the high extinction in this nebular region. 
This suggests that the region is extremely young, also affirmed by the fact that there is no free-free radio emission 
seen in the nebulous part (see Section \ref{section_RadioAnalysis}). 
There appears to be clustered star formation going on around the IRS\,9 
region. The IRS\,9N4 source from \citet{ojha04a} is resolved into two distinct sources in this image.  
The H$^{13}$CO$^{+}$($J$=1 - 0) molecular line emission peak lies to the eastern portion of the IRS\,9 sources, a 
region deficient in YSOs. This indicates the presence of dense gas in this region. A 450\,$\mu$m clump from \citet{reid05} 
is close to this peak.
Using the same procedure as for the IRS\,1--3 region (see Section \ref{section_morphology_IRS13}),  
a peak integrated line intensity of 1.5\,K\,km\,s$^{-1}$, and a source size at half intensity level of $\sim$ 0.77\,pc, 
LTE column density n(H$^{13}$CO$^{+}$) is calculated to be $\sim 2.7 \times 10^{12}$\,cm$^{-2}$. 
This leads to n(H$_2$) range of $\sim$ 0.9--5.4$\times$10$^{22}$\,cm$^{-2}$, and an LTE clump mass range of $\sim$ 100--660\msun\,.  
The spectrum for the clump is shown in Figure \ref{fig_H13CO+}\textit{(lower)}. The virial mass - following same steps 
as for IRS\,1--3 region - is approximately of the order of 500\msun\,. Alternatively, as Section \ref{section_morphology_IRS13},  
assuming gravitational equilibrium gives X(H$^{13}$CO$^{+}$) $\sim$ 0.65$\times$10$^{-10}$.

\subsection{Radio Continuum Emission}  
\label{section_RadioAnalysis} 

\begin{figure*}
\centering
\includegraphics[trim={1.5cm 5.0cm 0cm 5.5cm}, clip, scale=0.8]{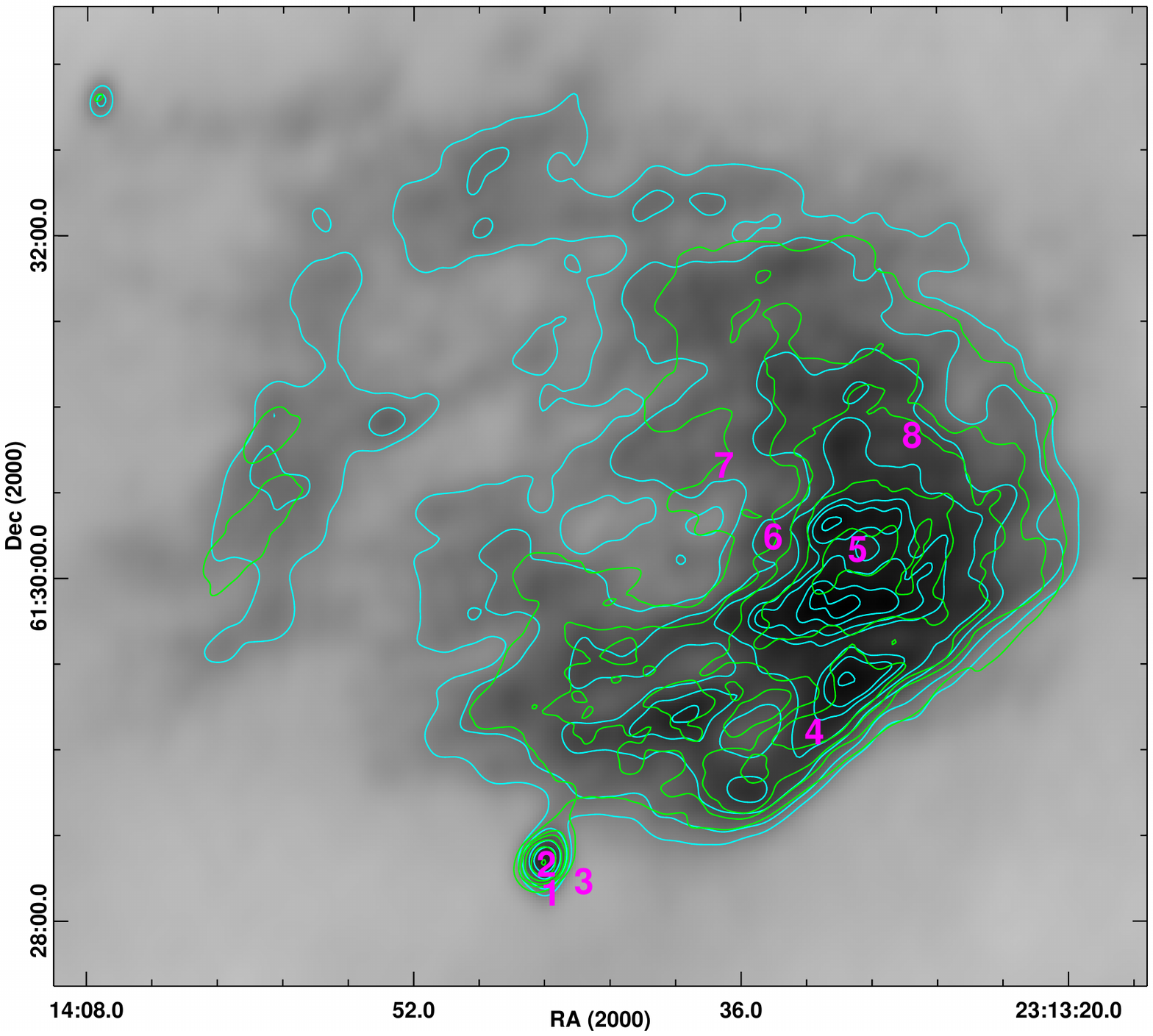}
\caption{The maximum resolution radio continuum images for 325\,MHz (cyan contours; 12.3\arcsec\,$\times$\,8.7\arcsec\,) and 
610\,MHz (green contours; 9.0\arcsec\,$\times$\,4.6\arcsec\,) obtained using GMRT for the entire NGC\,7538 region. The 
contours for 325 MHz image are 
drawn at 12, 17, 25, 30, 35, 37, 40, and 42 $\sigma$ ($\sigma \sim$ 1.78\,mJy). 610\,MHz contours are drawn at 
2.5, 5, 7, 10, 15, and 20 $\sigma$ ($\sigma \sim$ 4\,mJy). The IRS sources in this FoV are marked by their respective 
numbers in magenta. The background image is 325\,MHz radio image.} 
\label{fig_Radio_highres_325_610} 
\end{figure*}

\begin{figure*}
\centering
\subfigure
{
\includegraphics[trim={1.5cm 3.0cm 0cm 5cm}, clip, scale=0.43]{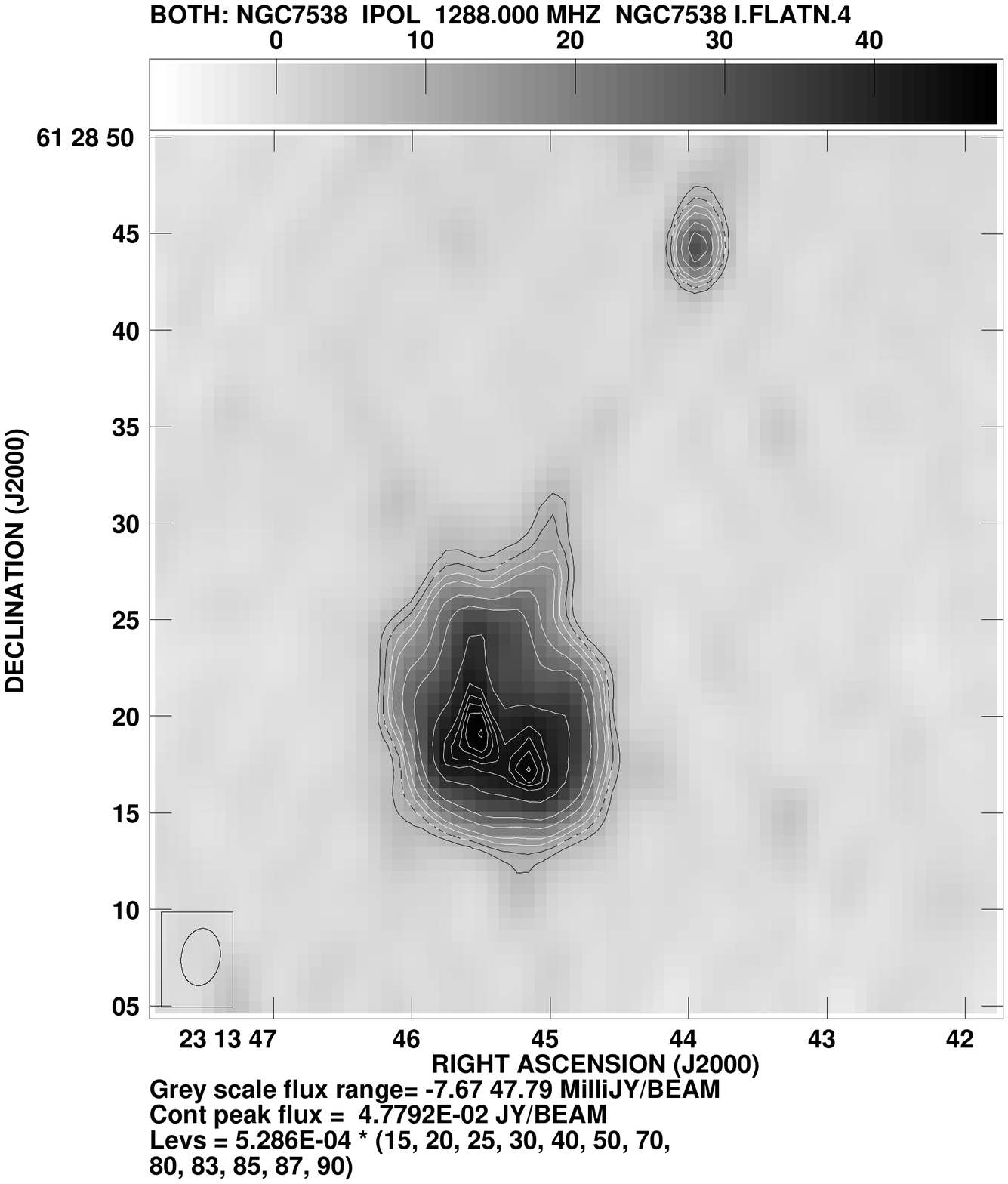}
% \label{fig_Radio_highres1280_IRS13} 
}
\subfigure
{
\includegraphics[trim={1.5cm 3.0cm 0cm 5cm}, clip, scale=0.43]{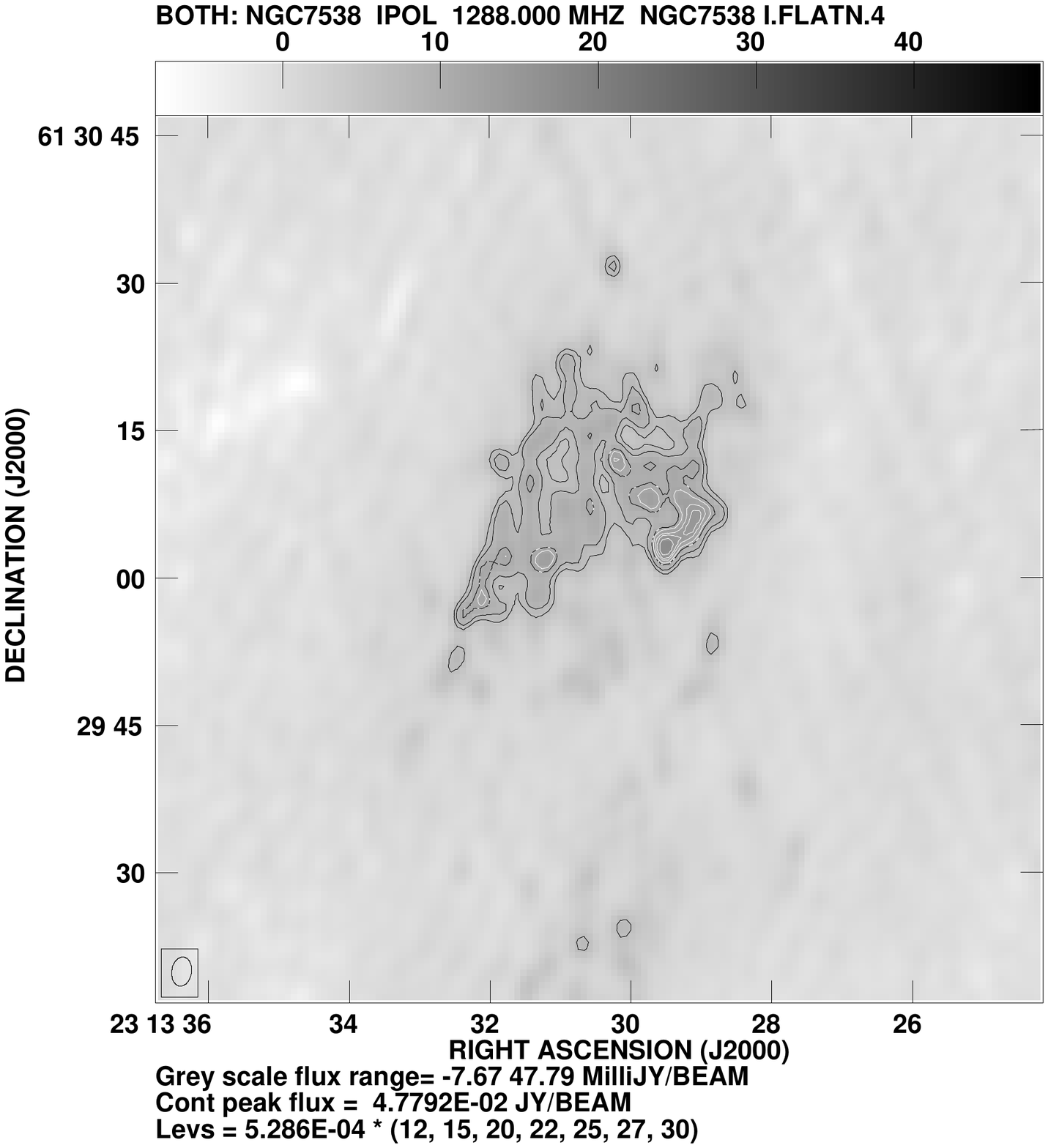}
% \label{fig_Radio_highres1280_IRS5} 
}
\caption{The maximum resolution radio continuum images for 1280\,MHz (3\arcsec\,$\times$2\arcsec\,). \textit{(left)} 
IRS\,1-3 region. Three separate cores are resolved here - to the North, East, and West. The contours are at 
15, 20, 25, 30, 40, 50, 70, 80, 83, 85, 87, and 90 $\sigma$ ($\sigma \sim$ 0.53\,mJy). \textit{(right)} High 
resolution image of the north-west part of the NGC\,7538 region, around the source marked 
\textquoteleft 5\textquoteright\, in Figure \ref{fig_Radio_highres_325_610}. The contours are drawn at 
12, 15, 20, 22, 25, 27, and 30 $\sigma$ ($\sigma \sim$ 0.53\,mJy).}   
\label{fig_Radio_highres_1280}  
\end{figure*}

The maximum resolution radio continuum images of the entire NGC\,7538 region which could be constructed are shown   
for 325\,MHz (resolution $\sim$ 12.3\arcsec\,$\times$\,8.7\arcsec\,), 610\,MHz 
(resolution $\sim$ 9.0\arcsec\,$\times$\,4.6\arcsec\,), and 1280\,MHz (resolution $\sim$ 3\arcsec\,$\times$\,2\arcsec\,) 
(Figures \ref{fig_Radio_highres_325_610} and \ref{fig_Radio_highres_1280}). 
The positions of the IRS sources in the region have been indicated by numbers in Figure \ref{fig_Radio_highres_325_610}. 
The radio contours in Figure \ref{fig_Radio_highres_325_610} show a definite champagne flow morphology 
\citep{tenorio79, whitworth79}. While the 
contours are closely packed in the south-west corner, they become spread as one moves from the south-west to the 
north-east direction. The south-west corner is density bounded, while the north-east side is ionization bounded. 
In the relatively lower resolution 325 and 610\,MHz maps (Figure \ref{fig_Radio_highres_325_610}), there are definite 
peaks associated with IRS\,1-3, as well as with the other IRS sources 4 and 5 (note that these IRS\,4 and 5 sources are 
outside our NIR FoV). The peak around the IRS\,1-3 region 
is resolved into three separate peaks - to the North, East, and West - in the higher resolution 1280\,MHz image 
(Figure \ref{fig_Radio_highres_1280}\textit{(left)}). Out of these, the IRS\,1-3-East peak coincides with the IRS\,2 source. 
As can be seen in this high-resolution image, the corresponding radio source for IRS\,2 has a cometary appearence, as 
has also been observed at higher radio frequencies \citep{bloomer98, campbell88}. 
The IRS\,1-3-North and West peaks do not have any NIR sources associated with them, though the IRS\,1-3-West peak (keeping in 
mind that the beam size is $\sim$ 3\arcsec\,$\times$\,2\arcsec\, here) is very close to the IRS\,2C peak from the 6\,cm 
radio map of \citet{campbell88}. The IRS\,1-3-East and West cores  
are approximately of the size of the synthesized beam ($\sim$3\arcsec\,$\times$2\arcsec\,), with a flux of $\sim$ 50\,mJy 
and 30\,mJy, respectively. The IRS\,1-3-North core is $\sim$3.8\arcsec\,$\times$2.3\arcsec\, in size, with a total 
flux of $\sim$43.4\,mJy. Based on the size of the cores \citep{kurtz02}, the whole IRS\,1-3 core can be classified as 
a compact \HII\, region (size from 0.1 to 0.5\,pc), while the North, East, and West cores can be classified as  
ultracompact \HII\, regions (size $\la$ 0.1\,pc). The high resolution 1280\,MHz image near the north-west 
part of the NGC\,7538 region (near the sources marked 4 and 5 in Figure \ref{fig_Radio_highres_325_610}; 
outside our NIR FoV) also shows multiple cores (Figure \ref{fig_Radio_highres_1280}\textit{(right)}). 

\begin{figure*}
\centering
\includegraphics[trim={1.5cm 7.0cm 0cm 8.5cm}, clip, scale=0.6]{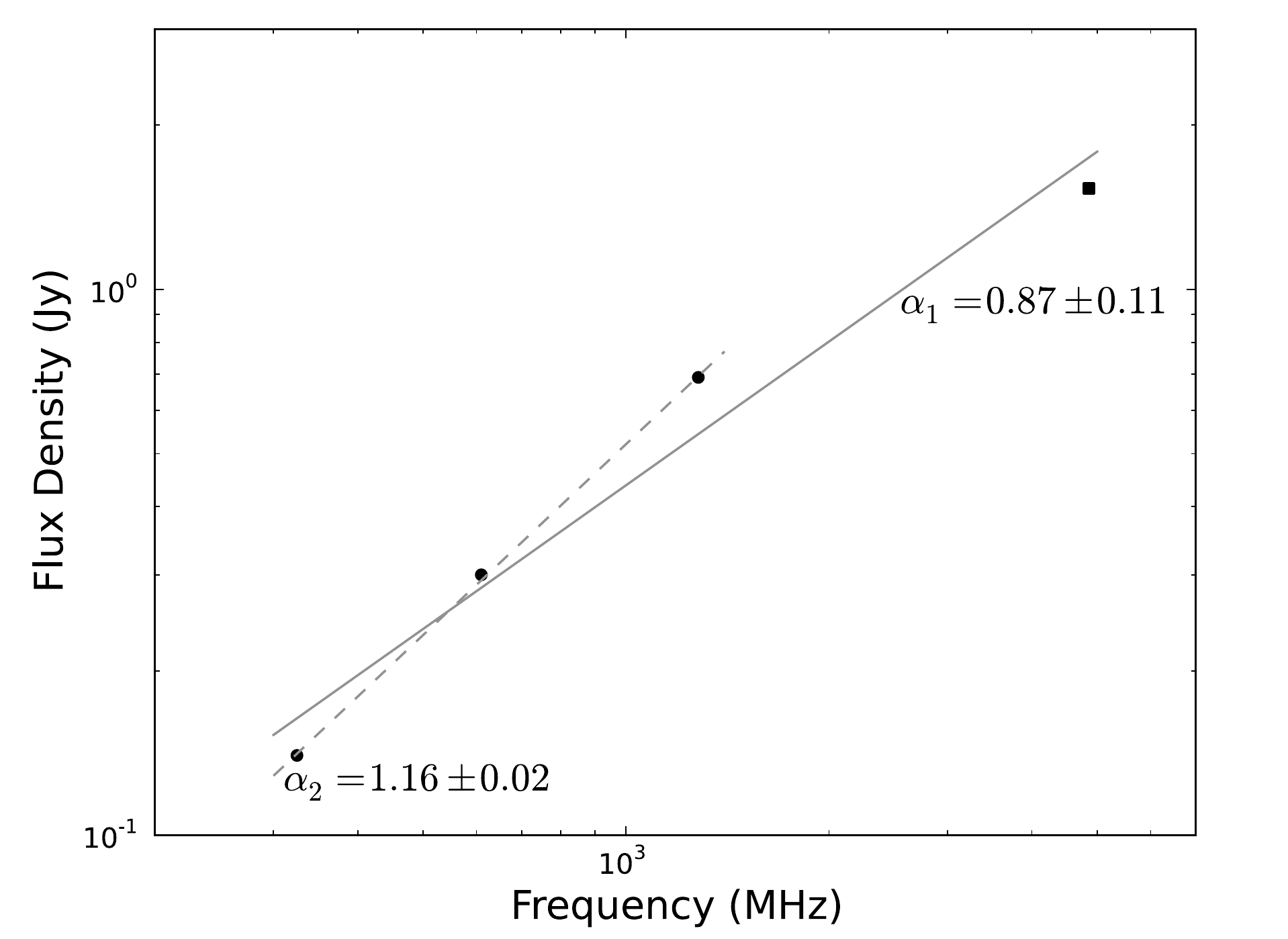}
\caption{SED fitting for the IRS\,1-3 compact \HII\, region. The circular markers show the data points from the GMRT at 
325, 610, and 1280\,MHz, while the square marker shows the VLA data point at 4860\,MHz. The solid grey line is the 
fitting using all four data points (with spectral index $\alpha _1 = $0.87$\pm$0.11), and the dashed grey line shows the 
fit using only the GMRT points (with spectral index $\alpha _2 = $1.16$\pm$0.02).} 
\label{fig_Radio_SED} 
\end{figure*}

The nature of a free-free emission region can be studied by calculating its spectral index $\alpha$, given by 
$S_\nu \propto \nu^{\alpha}$, where $\nu$ is the frequency and $S_\nu$ is the integrated flux density of the region  
at $\nu$. We can calculate the spectral index using $d\log S_\nu / d\log \nu$. As  
free-free emission changes from optically thick to optically thin, $\alpha$ varies from 2 to -0.1 \citep{panagia75, olnon75}. 
First of all, we obtained all the GMRT images at the same uniform resolution of about $\sim$12\arcsec\,$\times$9\arcsec\, 
(as this is the maximum resolution map we could make for 325\,MHz). Next, we calculated the 
integrated flux density within the IRS\,1-3 compact \HII\, region for all three frequencies. In addition to the 
GMRT frequencies, VLA archival image at the frequency of 4860\,MHz (at a very similar resolution of 
$\sim$14.9\arcsec\,$\times$11.5\arcsec\,) was also obtained and integrated flux density calculated for the compact 
\HII\, region. The values obtained are given in Table \ref{table_RadioObservation}. Figure \ref{fig_Radio_SED} shows the 
radio SED fit using these data points. The solid grey line shows the SED fit using all four data points 
($\alpha_1 = 0.87 \pm 0.11$), while the dashed grey line shows the fitting ($\alpha_2 = 1.16 \pm 0.02$) using only the 
GMRT points. These values of $\alpha$ suggest that the region is optically thick at these low frequencies, in consistency 
with earlier radio studies \citep{akabane92}. 

If we assume that this compact \HII\, region is homogeneous and spherically symmetric, then we can calculate the 
Lyman continuum photon luminosity (photon\,s$^{-1}$) using the following formula from 
\citet[see their Equation\,5]{moran83} : 
\begin{equation}
S_{*}=8\times10^{43}\left(\frac{S_{\nu}}{mJy}\right)\left(\frac{T_{e}}{10^{4}K}\right)^{-0.45}
\left(\frac{D}{kpc}\right)^{2}\left(\frac{\nu}{GHz}\right)^{0.1}
\end{equation}
where $S_{\nu}$ is the integrated flux density in mJy from the contour map, $D$ is the distance in kiloparsec, 
$T_{e}$ is the electron temperature, and $\nu$ is the frequency in GHz for which the luminosity is to be calculated. 
Now choosing the VLA data point of 4.860\,GHz (as it will lie in the most optically thin regime of the free-free 
emission SED), and $T_e$ a typical value of 10000\,K \citep{panagia75, olnon75}, $S_*$ was calculated to be 
$\sim$ 1.01$\times$10$^{48}$ 
(i.e. $\log S_* \sim 48.00$). According to the tabulated values of \citet{panagia73}, this much flux would correspond 
to a ZAMS O9 star ($\log S_* = 48.08$). Using various techniques, the spectral types of IRS\,1, 2, and 3 in the literature 
have been found to be consistent with O6 ZAMS \citep{willner76, pestolazzi04} (though \citet{franco04} give O8.5), 
O9.5 ZAMS \citep{campbelland84, bloomer98}, and a B-type star 
\citep{puga10}, respectively. However, various analyses have found extremely high extinction towards the IRS\,1 source. 
\citet{willner76} has concluded that the optical depth associated with IRS\,1 is high enough to absorb 99\% of UV photons. 
Similarly large extinction properties have been found by \citet{campbell88} and \citet{beuther12} in their study of this 
source. Hence, if we exclude IRS\,1, then our Lyman continuum luminosity is consistent with that for a O9.5 ZAMS source 
(IRS\,3 contribution, being a B-type source, will be much lesser comparatively).
% Slightly lower calculated value 
% ($\log S_* \sim 48.00$), in comparison to the tabulated value ($\log S_* \sim 48.08$), could be due to the absorption by 
% dust in the region. 
This high extinction could be the reason why IRS\,1 source is not seen in our high resolution 1280\,MHz map in 
Figure \ref{fig_Radio_highres_1280}\textit{(left)}. Additionally, this caveat
of dust absorption of radiation suggests that O9 ZAMS spectral type should be the lower limit and the actual spectral type 
could be earlier than this.

\section{Luminosity and Mass functions} 
\label{section_LFandMF}

We use the YSO catalogs from Section \ref{section_YSOselection} to generate the $K$-band luminosity function (KLF) of 
the IRS\,1-3 and IRS\,9 stellar clusters. $K$ band, as opposed to $H$ or $J$ band, is used as : the effects of extinction
are minimum, it probes sources upto much fainter luminosities, and the results can be compared to the existing literature. 

\subsection{IRS\,1-3 region}
\label{section_LFandMF_IRS13}

\subsubsection{$K-$band luminosity function}
\label{section_KLF_IRS13}

KLFs for the IRS\,1-3 region were generated using 2 sets of YSOs taken from Table \ref{table_YSOs_IRS13} : the  
first set containing a combination of 
\textquoteleft F+T+P+red-sources\textquoteright\, (basically the entire Table
\ref{table_YSOs_IRS13}), and a second set containing only 
\textquoteleft T+P+Any source with X-ray detection\textquoteright\, 
(see Section \ref{section_YSOselection} and Figures \ref{fig_NIRCCD} and \ref{fig_NIRCMD}). This was done as 
the first set may contain YSO candidates with field star contamination and will need to be corrected for it, 
while the second set is most likely YSOs with much lower field contamination. 
To derive the KLF of a region, one needs to apply corrections for the following : star count 
incompleteness (which is a function of magnitude), and the field star contamination towards the cluster. 

Star count incompleteness was corrected for by using the completeness calculations from Section \ref{section_NIRreduction}. 
The completeness fraction had been obtained for each 0.5 magnitude bin. The counts in each magnitude bin were scaled up 
by dividing the counts by the completeness fraction in the respective bins. This gives us the completeness-corrected KLF. 

\begin{figure*}
\centering
\includegraphics[trim={7cm 4.0cm 9cm 5.5cm}, clip, scale=0.8, angle=90]{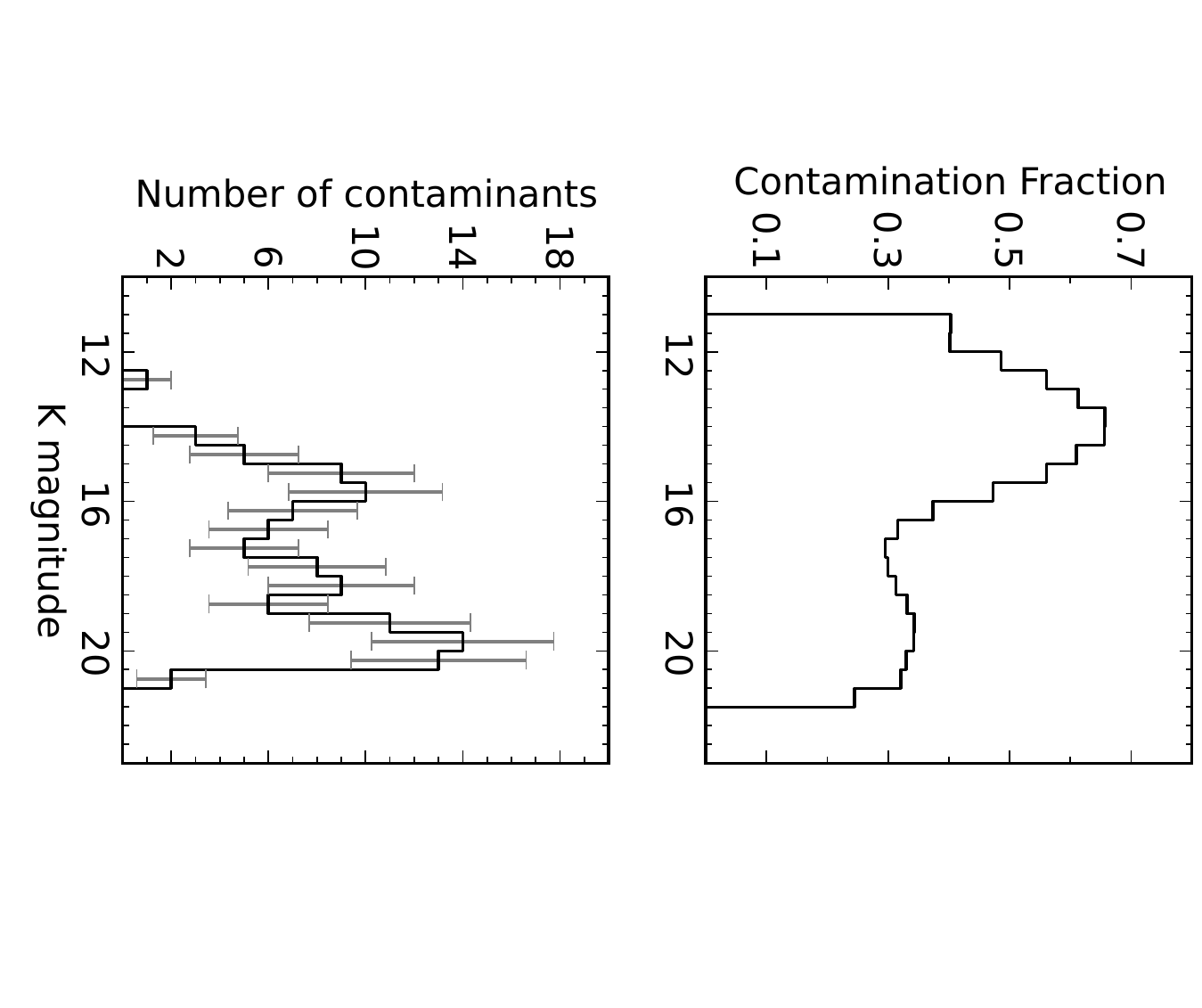}
\caption{\textit{(upper)} The contamination fraction (per magnitude bin) calculated using the Galactic model 
of \citet{robin03}. \textit{(lower)} The number of contaminants (per magnitude bin) for the IRS\,1-3 cluster region. The 
error bars show the error due to counting statistics. A similar exercise was carried out for the IRS\,9 region.} 
\label{fig_fieldstar_IRS13} 
\end{figure*}

The field star contamination was assessed using the sky field source catalog (from Section \ref{section_NIRreduction}) 
in conjunction with the Galactic model of \citet{robin03}, similar to \citet{ojha04a,ojha04b}. 
Two sets of catalogs were generated using the Besan\c{c}on model 
of stellar population synthesis \citep{robin03} in the direction of the sky field region. The first set contains the 
sources generated by setting $A_V = 4.4$\,mag, which was chosen as it is the average $A_V$ of sources in the sky field  
CM-D (between $H-K=0$ and the low density gap at $H-K=0.5$), and thus most likely to be the foreground extinction of 
sources towards this region. 
From this first catalog set, we get the total number of sources as well as the number of 
foreground sources in the direction of the sky field region. The foreground sources, in this catalog, 
are simply the ones with distances $< 2.65$\,kpc.
In addition to the total and 
foreground sources towards this region, we need to assess the background field star contamination too. 
However, the 
background field star contaminants will suffer an extra extinction due to the intervening molecular cloud, which needs to 
be taken into account. Now, the average extinction towards the NGC\,7538 region has been found to be $A_V = 15$ mag 
\citep{ojha04a}.    
If we assume a spherical geometry of the molecular cloud medium, then it follows that the sources behind the molecular 
cloud should suffer an extinction of $A_V = 15+15 = 30$\,mag. Therefore, we generated a second model catalog set 
by setting $A_V=30$\,mag.  
In this second set, all the sources with distances $> 2.65$\,kpc will give us the background contaminants. After having 
obtained the total, foreground, and the background sources, the $K$ band histogram (with binwidth$=$0.5) was plotted and 
the contamination fraction (foreground+background/total) was obtained for each bin. Finally, to obtain the absolute number of 
contaminating sources in each magnitude bin, we use the sky field $K$ band histogram. Each magnitude bin of the sky 
field $K$ magnitude histogram is scaled by the ratio ($\sim 0.75$) of the areas of sky field to IRS\,1-3 region, as well 
as by its contamination fraction calculated above. 
Fig. \ref{fig_fieldstar_IRS13} shows the contamination fraction, along with the number of field contaminants in  
the IRS\,1-3 region, as a function of magnitude. The obtained contaminant number in each 
bin was subtracted from the completeness-corrected KLF (binwise) to obtain the field- and completeness-corrected KLF. 

\begin{figure*}
\centering
\subfigure
{
\includegraphics[trim={2.5cm 7.0cm 3cm 5.5cm}, clip, scale=0.5]{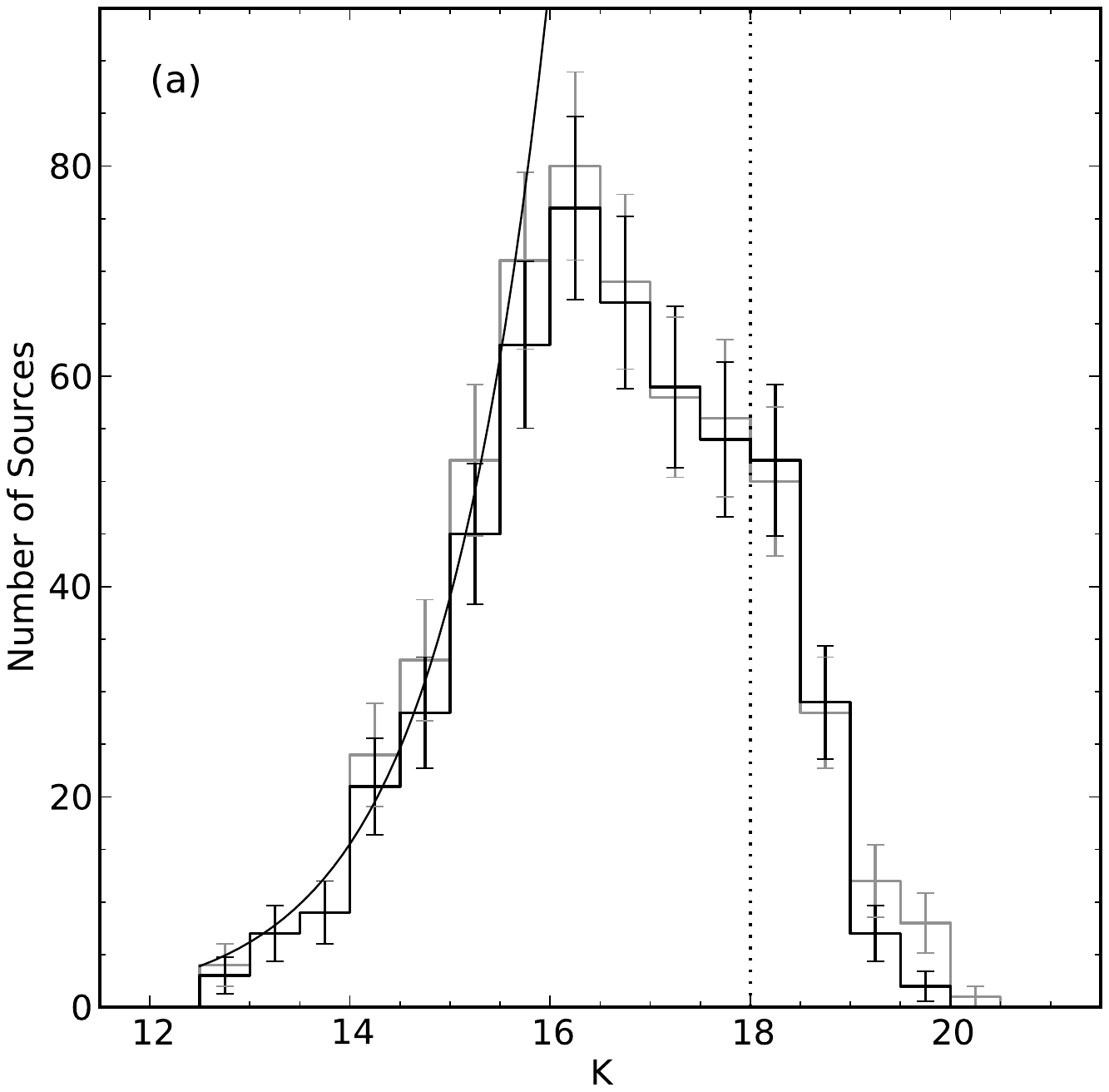}
\label{fig_KLF_IRS13_hist_cIIaIII}
}
\subfigure
{
\includegraphics[trim={2.5cm 7.0cm 3cm 7.5cm}, clip, scale=0.5]{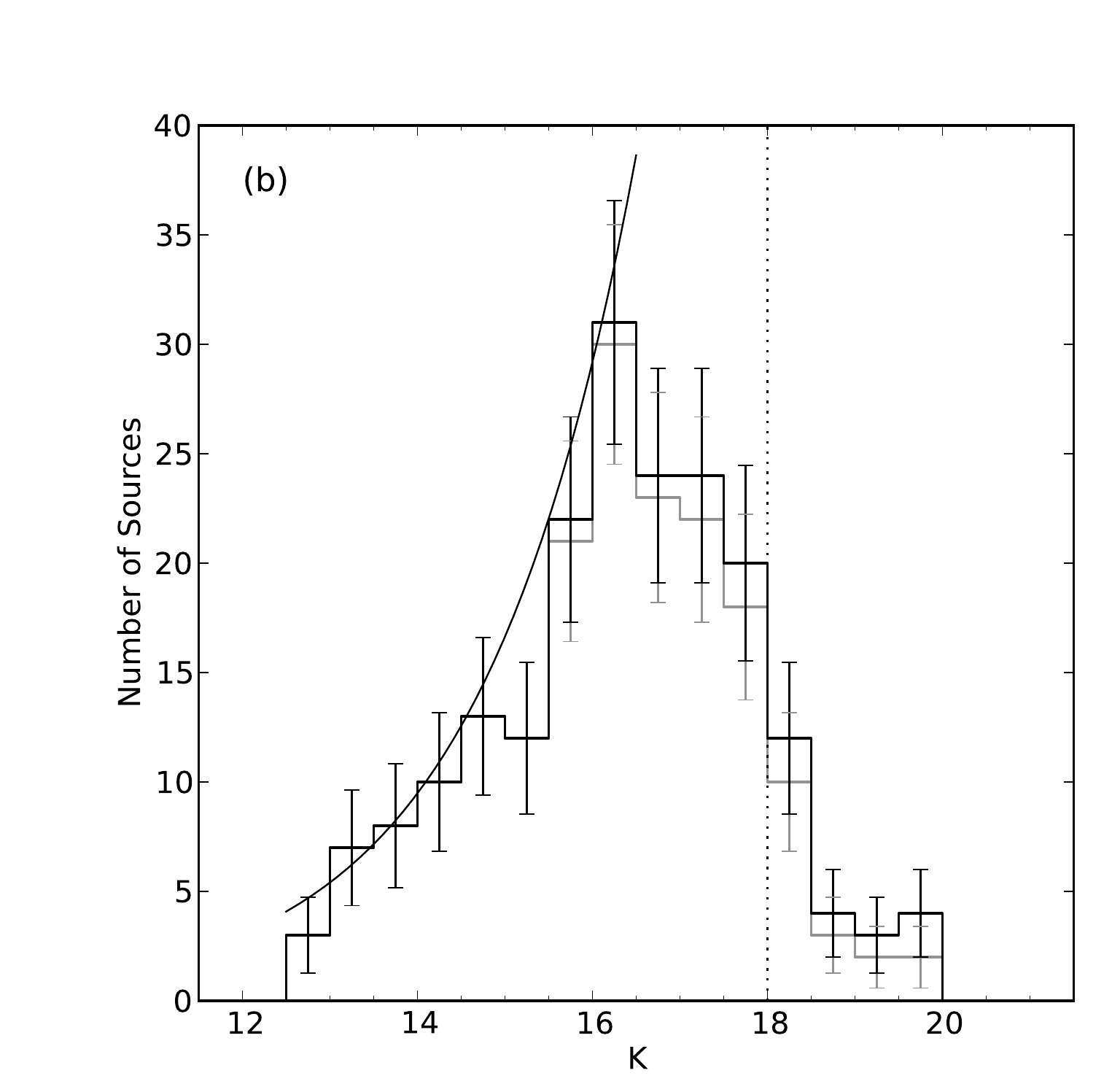}
\label{fig_KLF_IRS13_hist_cII} 
}
\caption{KLFs for the IRS\,1-3 region. 
(a) For \textquoteleft F+T+P+red-sources\textquoteright\, sources. Grey line histogram 
represents the raw KLF. Solid black line histogram shows the final field- and completeness-corrected KLF. 
(b) For \textquoteleft T+P+Any source with X-ray detection\textquoteright\, sources. Grey 
line is for the raw KLF, while the black line shows the final completeness-corrected KLF. The vertical 
dotted line in both figures marks the $K$ band 90\% completeness limit at 18 mag. The error in each bin is the  
Poissonian $\pm \sqrt{N}$ error. The black curve shows the power law fit to the KLF.}   
\label{fig_KLF_IRS13_hist} 
\end{figure*}

The KLFs for the first YSO set (\textquoteleft F+T+P+red-sources\textquoteright\,), and 
second YSO set (\textquoteleft T+P+Any source with X-ray detection\textquoteright\,) are shown in 
Figure \ref{fig_KLF_IRS13_hist}. A binsize of 0.5 has been chosen as it is much larger than the errors in source 
magnitudes. To compare with the literature, we calculate the $(d\log N/dm_{K})$ slope for the rising 
part of the KLFs to be $0.40 \pm 0.03$ (in the magnitude range 12.5--16.5) for the first YSO set, 
and $0.24 \pm 0.03$ (in the magnitude range 12.5--16.5) for the second YSO set. 
While the KLF slope for the first YSO set is higher than that for the completeness- and field-corrected KLF for the 
whole NGC\,7538 region ($0.30 \pm 0.03$) from \citet{ojha04a}, it is similar to that of the NGC\,1893 
($\sim$ 3.25\,kpc; $0.34 \pm 0.07$) from \citet{sharma07}, Tr\,14 ($\sim$ 2.5\,kpc; $0.37 \pm 0.01$) from \citet{sanchawala07},  
% Other young clusters with such a high value of this KLF slope are NGC\,1624 ($\sim$ 6.0\,kpc) \citep{jose11}, 
and IRAS\,06055+2039 ($\sim$ 2.6\,kpc; $0.43 \pm 0.09$) \citep{tej06}. 
% and NGC\,2264 cluster from \citet[see its Figure 4(a)]{lada93}; 
The slope for the second set is consistent with that calculated earlier in \citet{ojha04a} for the completeness-corrected 
whole NGC\,7538 region 
($0.28 \pm 0.02$) as well as that for the younger regions in NGC\,7538 ($0.27 \pm 0.03$). 
It is also very close to the value for the whole W3 region 
($\sim$ 1.83\,kpc; $0.26 \pm 0.01$) \citep{ojha04b} and Sh2-255 IR region ($\sim$ 2.6\,kpc; $0.17 \pm 0.03$) from \citet{ojha11}. 
A turnoff is seen in both the KLFs after 16-16.5 magnitude bin, similar to the Tr\,14 region from \citet{sanchawala07}.

\subsubsection{Mass Function} 
\label{section_MF_IRS13} 

The mass function (MF) is usually described by the following differential form :  
\begin{equation}
\label{equation_MF1} 
\xi(\log\,M_*) = dN/d\log M_* 
\end{equation}
where $N$ is the number of stars and $M_*$ is mass of stars \citep{bastian10, dantona98, chabrier03, kroupa13}. 
Since magnitude ($m_{K}$), and not mass ($M_*$), is an observable quantity, $\xi$ has to be re-written as : 
\begin{equation}
\label{equation_MF2}
\xi(\log\,M_*) = dN/d\log M_* = (dN/dm_{K}) \div (d\log M_*/dm_{K})  
\end{equation}
where $(dN/dm_{K})$ is nothing but the KLF slope, and $(d\log M_*/dm_{K})$ represents the derivative of 
the so-called mass-luminosity relation (MLR). If we have the form of the KLF and the MLR derivative, 
then $\xi$ can be evaluated. 

\begin{figure*}
\centering 
\includegraphics[trim={1.5cm 6.8cm 0cm 8cm}, clip, scale=0.7]{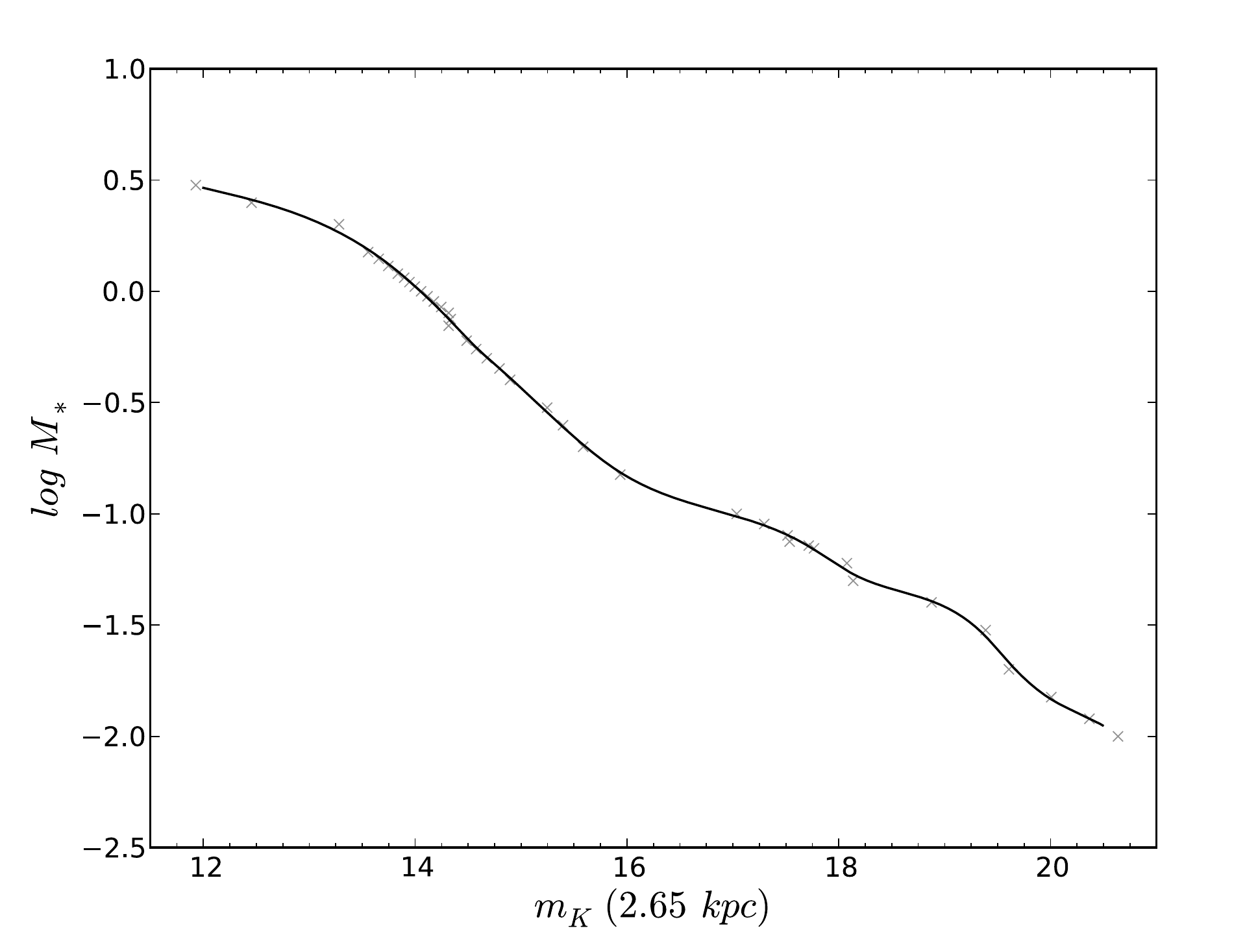}
\caption{\textit{(upper)} MLR from literature (see text). The grey crosses mark the 
data points from literature and the black line shows the fitted curve to it. 
We confine our analysis to $m_K \geq 12$, i.e. lower limit of MLR.}   
\label{fig_MLR} 
\end{figure*}

The MLR depends on the age of the cluster, and in our case we use the age estimate of 1\,Myr for the NGC\,7538 
region from \citet{ojha04a}. The theoretical isochrones (for 1\,Myr) from \citet{baraffe03} (for the mass range 
$0.01-0.1$ M$_{\odot}$), \citet{baraffe98} 
($0.15-1.50$ M$_{\odot}$), and \citet{palla99} ($2.0-3.0$ M$_{\odot}$) are used for the MLR. 
The absolute magnitudes were converted to apparent magnitudes using the distance modulus (at a 
distance of 2.65\,kpc). The complete MLR is shown in Figure \ref{fig_MLR}, along with the curve 
fit to it. To calculate $\xi$, we need  
the derivative of this MLR. We limit our analysis to sources with $m_K \geq 12$ mag,  
as it the lower limit of the MLR.  

In addition to the MLR, a form of the intrinsic KLF is also needed. To derive the intrinsic KLF, the sources 
were corrected for extinction by dereddening them along the reddening vectors to the respective loci as per 
the following order of steps. First, the sources which were present in the \textquoteleft F\textquoteright\, 
region of the NIR CC-D (Figure \ref{fig_NIRCCD_IRS13}) were dereddened to the dwarf locus \citep[whose low-mass 
regime - from the turn-over onwards - was approximated 
by a straight line; similar to][]{tej06, samal07}, while those in the \textquoteleft T\textquoteright\, and 
\textquoteleft P\textquoteright\, regions were dereddened to the CTTS locus. 
The rest of the sources (those not 
dereddened using the NIR CC-D) were dereddened using the $H$ and $K$ magnitudes to the ZAMS locus (approximated 
by a straight line) in the NIR CM-D (Figure \ref{fig_NIRCMD_IRS13})   
to estimate their extinctions, though this is only an approximation as the YSOs are most-likely not on ZAMS yet.
The major source of uncertainty in using the CM-D is the correction of the locus for distance, which need not be 
well-determined. 
The individual visual extinction of the sources was found to range upto 60\,mag. This is much deeper, as expected, 
than upto 40\,mag which was estimated by \citet{ojha04a} for their detected YSOs. 
The histogram (not shown here) of these visual extinctions was found to peak in the range 7.5-10\,mag. 
The mean extinction value was $\sim$\,15\,mag, consistent with that used for model simulations in 
Section \ref{section_KLF_IRS13} and \citet{ojha04a}. The sources 
with $A_V > 40$\,mag were mostly found to be distributed along the southern boundary between the nebulous and non-nebulous 
region (also see Figs. \ref{fig_ColourComposite_ForCompleteness}\textit{(left)} and \ref{fig_Morphology_IRS13}). 
A few such sources were also associated with the small nebular patch at the south-east corner of the 
IRS\,1-3 region. 

Using this catalog of reddening-corrected sources, field-, completeness-, and reddening-corrected KLFs were obtained 
for the two YSO sets (\textquoteleft F+T+P+red-sources\textquoteright\, being the first set 
and \textquoteleft T+P+Any source with X-ray detection\textquoteright\, being the second set; 
see Section \ref{section_KLF_IRS13}). To preserve the information about the form of the KLF, different magnitude intervals
(each interval containing three or more bins) were fit with equations of straight line using simple linear regression. 
This gives us the KLF slope for each magnitude bin.   

\begin{figure*}
\centering 
\includegraphics[trim={0cm 6.0cm 0cm 6.5cm}, clip, scale=0.5]{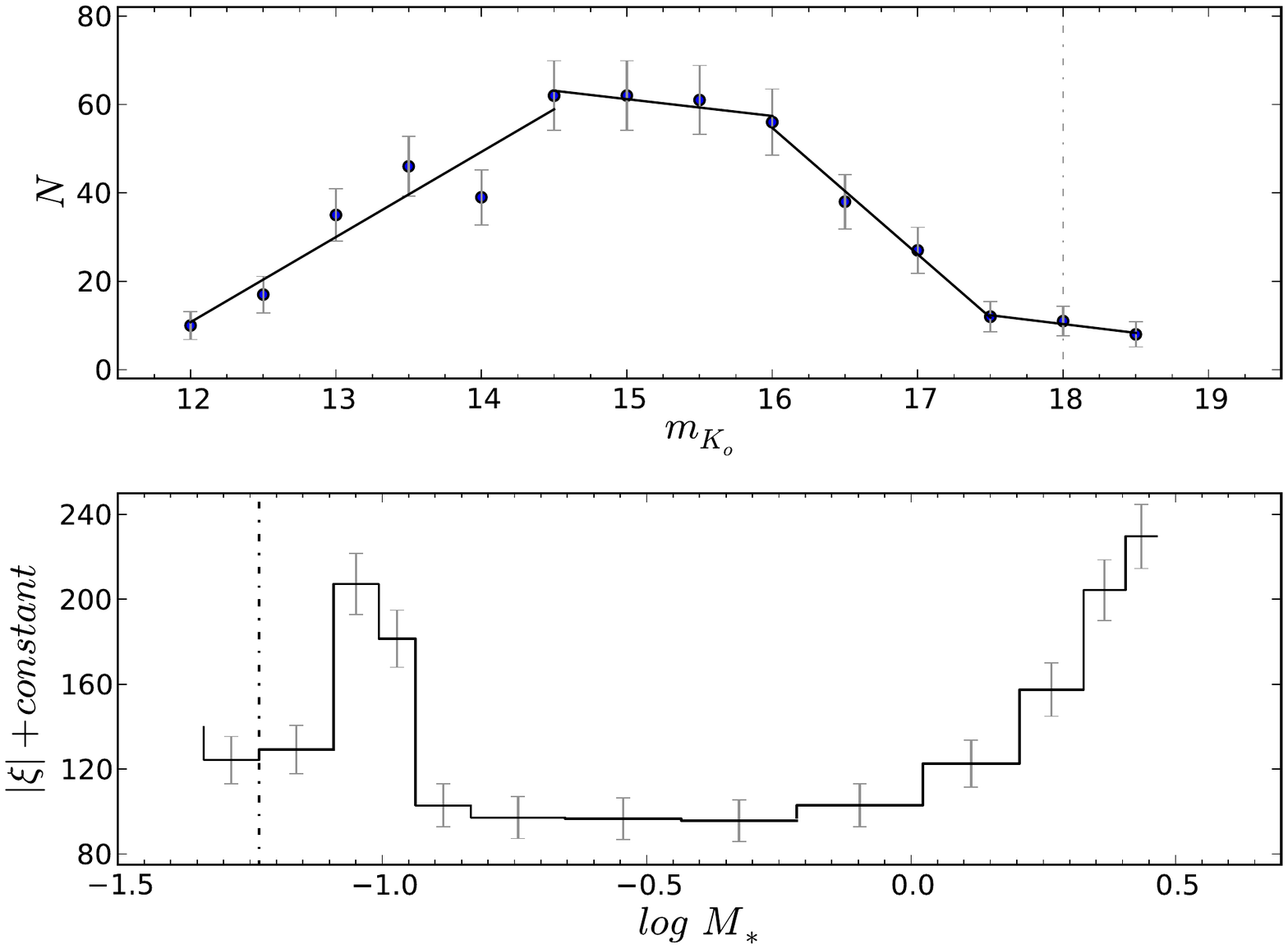}
\caption{KLF fits and $\xi$ (differential form) for 
\textquoteleft F+T+P+red-sources\textquoteright\, 
YSO catalog in the IRS\,1-3 region. \textit{(upper)} The field-, completeness-, and reddening-corrected KLF and the 
straight line fits to its different portions. 
\textit{(lower)} $\xi$ calculated using equation \ref{equation_MF2}. Each 0.5 mag bin from the KLF has been mapped to 
a corresponding bin in $\log M_*$ space here. The vertical dashed-dotted line indicates the 90\% completeness limit. Poissonian 
errors are marked for each bin.}   
\label{fig_KLFfit_MF_cIIaIII}
\end{figure*}

\begin{figure*}
\centering 
\includegraphics[trim={0cm 6.0cm 0cm 6.5cm}, clip, scale=0.5]{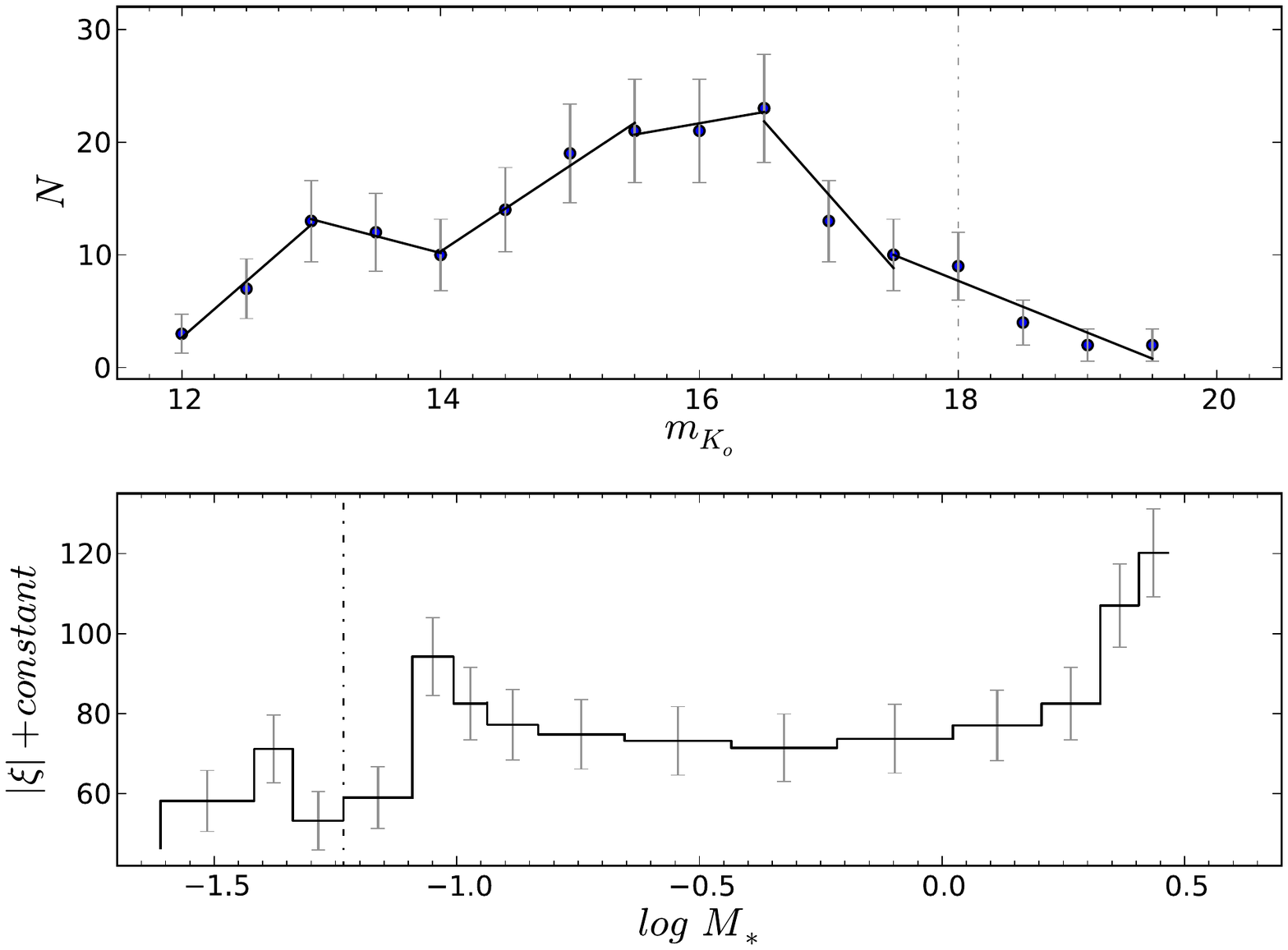}
\caption{KLF fits and $\xi$ for \textquoteleft T+P+Any source with X-ray detection\textquoteright\, 
YSO catalog in the IRS\,1-3 region. Rest is same as for Figure \ref{fig_KLFfit_MF_cIIaIII}.}   
\label{fig_KLFfit_MF_cII}
\end{figure*}

After we obtain the KLF slope $(dN/dm_{K})$ in each magnitude bin and the form of the MLR 
derivative ($d\log M_*/dm_{K}$ at discrete points), value of $\xi$ is calculated using Equation \ref{equation_MF2} 
for each magntiude bin - which is further mapped onto the $\log M_*$ space using the MLR. 
The resulting form of $\xi$ is shown in 
Figures \ref{fig_KLFfit_MF_cIIaIII} and \ref{fig_KLFfit_MF_cII} along with the KLFs. Poissonian error of $\pm \sqrt{N}$ 
is marked for each bin. It should be kept in mind that equal magnitude intervals do not map onto equal $\log M_*$ intervals. 
The shape of the $\xi$ is closest to that derived by \citet{scalo86} for field stars, with a peak at the 
low mass end, and another peak at the intermediate mass \citep{meyer00}. The low mass peak is at $\log M_*$ bin 
of $-1 .09$ to $-1.00$, i.e $0.08-0.1\,M_\odot$ (for Figures \ref{fig_KLFfit_MF_cIIaIII} and \ref{fig_KLFfit_MF_cII}). 
This rise in the MF till the BD limit has been seen for other regions like W3 Main \citep{ojha09} and S106 \citep{oasa06} 
too, as well as mentioned likely by \citet{kroupa07}. The curve does not extend enough in the intermediate mass range to 
clearly discern the peak. 

\begin{figure*}
\centering 
\includegraphics[trim={1.5cm 9cm 0cm 10.5cm}, clip, scale=0.6]{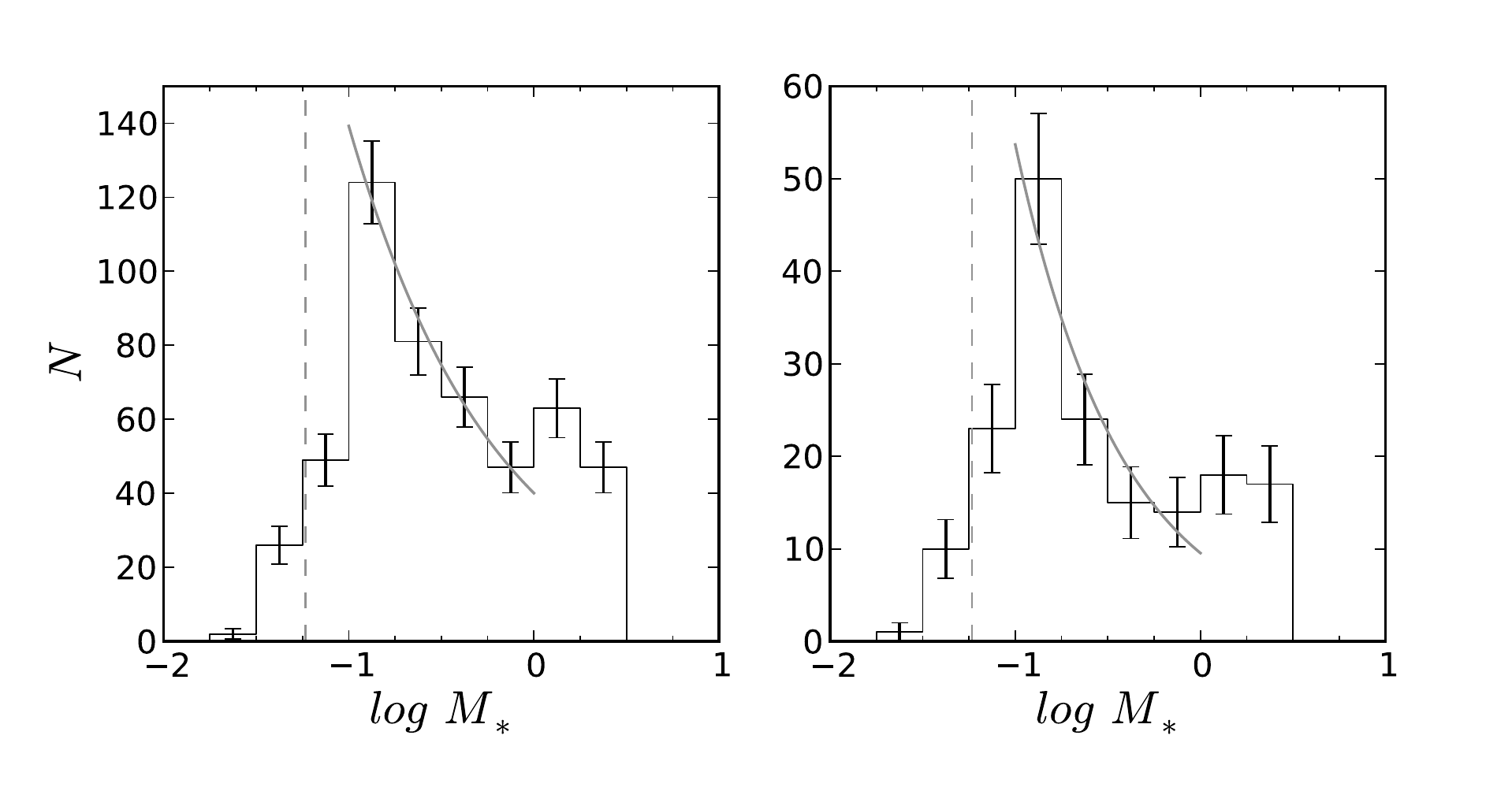}
\caption{IRS\,1-3 region : 
$N(\log M_*)$ for \textit{(left)} \textquoteleft F+T+P+red-sources\textquoteright\, 
YSO catalog, and \textit{(right)} \textquoteleft T+P+Any source with X-ray detection\textquoteright\,
YSO catalog obtained by assigning mass to each source individually and binning. Error bars in each bin show Poissonian 
error. The grey curve shows the fit to the MF in 0.1--1 M$_\odot$ mass range.  
The vertical dashed line indicates the mass corresponding to the $K$ 90\% completeness limit of 18\,mag. The binwidth is 0.25.}   
\label{fig_MF_histogram}
\end{figure*}

Another way to test the MF form is to use the MLR to assign a mass to each star in the catalog, and then bin those 
masses to obtain the $N(\log M_*)$, rather than the differential form $\xi(\log M_*)$  \citep[see its Section 1.3]{chabrier03}. 
Using this, we obtain the MFs shown in Figure \ref{fig_MF_histogram}. The field star subtraction for the 
first catalog here was carried out statistically using the NIR CM-D as follows. 
Similar to \citet{sharma07}, we divided the CM-Ds of the IRS\,1-3 region and the sky field region into
grids with $\Delta K$=0.5\,mag and $\Delta (H-K)$=0.1\,mag. The number of stars was compared on a grid-by-grid basis, and 
from the 
IRS\,1-3 CM-D, a fixed number of stars (equal to the number in the corresponding sky field grid) was removed based on 
their distances to the sky field stars in the colour-magnitude space. Each grid was corrected for incompleteness.

As can be seen, the MFs are similar in shape to those in Figures \ref{fig_KLFfit_MF_cIIaIII} and \ref{fig_KLFfit_MF_cII}, 
with a peak in the $\log M_* =$ -0.75 -- -1.0 bin (i.e. 0.1--0.18\,M$_{\odot}$), and another at 
$\log M_* =$ 0--0.25 bin (i.e. 1--1.78\,M$_{\odot}$). 
However, the peak at the lower mass regime, though consistent with the other method, is on the slightly higher side here. 
The mass range of the turn-off point in the lower mass regime is consistent with those for other prominent clusters in the 
literature 
\citep{bastian10}, like $\sigma$\,Orionis \citep{pena12}, $\rho$\,Ophiuchi \citep{alves12}, IC\,348 \citep{alves13a}, and 
Orion Nebula Cluster \citep{hillenbrand00}. A secondary peak seen here is also observed in the IMF of the $\rho$\,Ophiuchi 
cluster from \citet{alves12} and Orion Nebula Cluster from \citet{hillenbrand00}. 
If we assume a power law form of the MF \citep[similar to][]{salpeter55}, then : 
\begin{eqnarray}
dN/d(\log M_*) & \propto & M_*^{- \Gamma} \\
\Rightarrow dN/dM_* & \propto & M_*^{- (\Gamma + 1)}. 
\end{eqnarray} 
Now, if we additionally assume that the star formation is strictly coeval, then it can be mathematically shown that 
\begin{equation}
\label{equation_IMFGamma}
- \Gamma = d(\log N)/d(\log M_*), 
\end{equation} 
and that the present day mass function will have the same slope as the IMF \citep[see its Section\,2.1]{massey98}. 
Also, since the age of this region is $\sim$ 1\,Myr, which is lower than the MS life of even the most massive 
stars, none of the stars would have disappeared from the field. In Figure \ref{fig_MF_histogram}, fitting the sub-solar 
low mass range (between the first and the second peaks, i.e. 0.1-1\,M$_\odot$) using Equation \ref{equation_IMFGamma},  
we get the value of $\Gamma$ for the first and second YSO sets as $0.54 \pm 0.05$ (say, $\Gamma_{1}$) and $0.75 \pm 0.18$
(say, $\Gamma_{2}$), respectively. Both the slopes are lower than the Salpeter slope of 1.35. While $\Gamma_1$ seems 
consistent with that from \citet{kroupa02} (also see Figure 2 of review by \citealt*[]{bastian10}), $\Gamma_2$ is slightly 
steeper. This steepness, though, might be explained by the fact that Figure \ref{fig_MF_histogram}\textit{(right)} includes 
mostly the youngest sources with fainter magnitudes - thus making this MF \textquoteleft bottom-heavy\textquoteright\,, 
i.e. more sources at lower mass ranges. On the other hand, Figure \ref{fig_MF_histogram}\textit{(left)} includes 
Class\,III sources and thus more sources in relatively higher mass ranges. In general, more complicated/realistic forms of 
the IMF can emerge due to accretion processes, leading to a tail towards the high mass end \citep{dib10}, which will 
become noticeable only in high-mass star-forming regions where the intermediate to high-mass bins are well populated. 
Finally, we should keep in mind the caveat that any derivation of an IMF suffers from multiple unavoidable and systematic 
biases, e.g. those 
arising due to - among others - choice of the stellar MLR, PMS evolution, effect of unresolved sources, binning, etc. 
\citet[see their Section 2.1]{kroupa13} have dealt with these biases in a succinct manner. Different treatment of these 
biases might accordingly alter a derived IMF. 

The ratio of stars to BDs in a region is often used as another quantitative indicator for the mass function. We 
consider the following two definitions of this ratio : 
\begin{eqnarray}
R_1 &=& N(0.08 < M/M_{\odot} \leq 1)/N(0.03 \leq M/M_{\odot} \leq 0.08), and \\
R_2 &=& N(0.08 < M/M_{\odot} \leq 10)/N(0.02 \leq M/M_{\odot} \leq 0.08),  
\end{eqnarray}
similar to \citet{scholz12}. However, since our observations are complete only upto 0.06\,M$_\odot$, the R$_1$ and R$_2$ 
values obtained will be the upper limit. Taking this caveat into account and using the above equations, we get the value of 
$R_1 < 349/44 \sim 8$, and $R_2 < 470/46 \sim 10.2$. The compilation by \citet{scholz12} shows that the values of $R_1$ and $R_2$ 
for other star-forming regions lie in the range 2--8, while \citet{luhman07} state that $R_2$ ranges from 5--8 in star-forming 
regions. On the other hand, \citet{alves12} find higher value of $R_2$ (upto 11 depending on their analysis parameters) in the 
$\rho$\,Ophiuchi cluster. However, given the fact that the values of $R_1$ and $R_2$ calculated here are upper limits, they 
are consistent with those of other star-forming regions.

\subsection{IRS\,9 region}
\label{section_KLF_IRS9}

\begin{figure*}
\centering
\includegraphics[trim={1.5cm 9.0cm 0cm 8.5cm}, clip, scale=0.8]{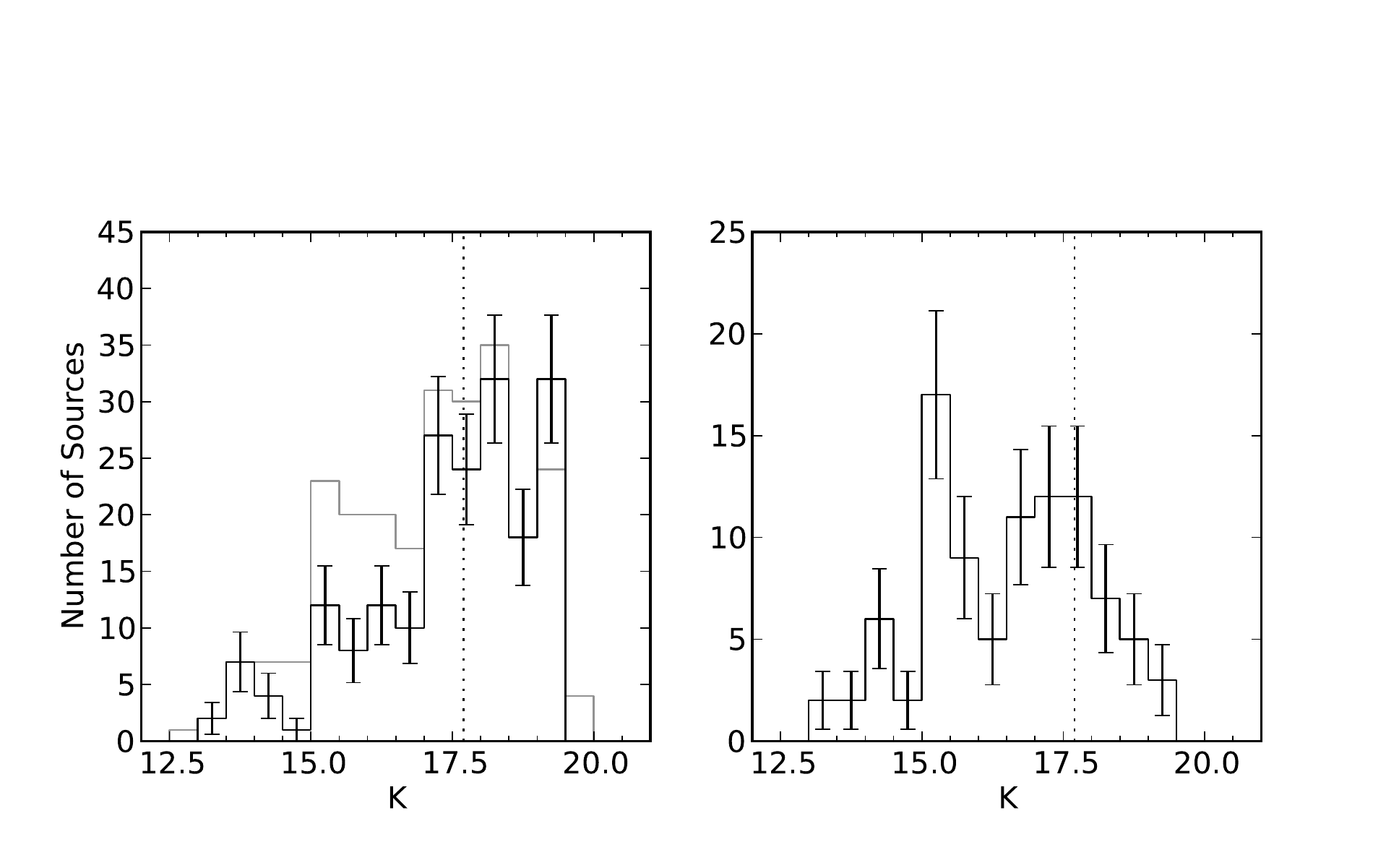}
\caption{KLFs for the IRS\,9 region. \textit{(left)} The grey line histogram shows the raw KLF of all sources 
detected in $K$-band, while the black line shows the KLF after field- and completeness-correction. 
\textit{(right)} KLF of sources which had $H-K>1$, or were detected in X-ray. Error bars are Poissonian. 
The vertical dotted line marks the $K$ band 90\% completeness limit at 17.7 mag.} 
\label{fig_KLF_IRS9_hist} 
\end{figure*}

We obtained the KLF for the IRS\,9 region using the catalog of sources identified in at least $K$ band, as well as the 
catalog of candidate YSOs (red sources with $H-K>1$ identified in Section \ref{section_CMD} and those with X-ray emission). 
The KLFs are shown in Figure \ref{fig_KLF_IRS9_hist}. Figure \ref{fig_KLF_IRS9_hist}\textit{(left)} shows the raw KLF 
(grey line) along with the field- and completeness-corrected KLF (black line). Since the statistics are low for this 
region, we only make a few qualitative comparisons here. 
The $(d\log N/dm_{K})$ slope was calculated 
to be $0.20 \pm 0.04$ (in 13--19 $K$ mag range), and is lower than that calculated for younger regions in 
\citet{ojha04a} and some other star-forming regions (see Section \ref{section_KLF_IRS13}), though similar to 
Sh2-255\,IR region \citep{ojha11}.  
The KLF of red sources ($H-K>1$) is shown in Figure \ref{fig_KLF_IRS9_hist}\textit{(right)}. 
There appear to be two peaks in Figure \ref{fig_KLF_IRS9_hist}\textit{(right)} - 
one near 15--15.5 mag and another at about the completeness limit (marked by a dotted vertical line). The KLF seems to be 
plateauing near the completeness limit.   
The brighter peak is most likely to be due to field contamination, which has not been corrected for due to a lack of statistics
here. If we consider the 16--17.7 mag interval of this YSOs' KLF, we see that it first has a steeper slope in the 
16--16.5 mag interval, 
and then plateaus off in the 16.5--17.7 mag interval. This steep slope followed by plateauing is seen in other regions like 
NGC\,1624 \citep{jose11} and Sh2-255\,IR region \citep{ojha11} too, albeit for different magnitude limits. In this range (16--17.5), 
the slope comes out to be $0.24 \pm 0.11$, which is very similar to that from \citet{ojha04a} for younger regions of 
NGC\,7538, and is consistent within errors with that obtained for the Sh2-255\,IR region \citep{ojha11}.

\section{Discussion}
\label{section_discussion} 

Analysis of star-forming regions in different stages is essential in understanding how star formation proceeds 
and the effect of manifold physical processes on various diagnostic tools (like LF, MF, etc). In general, 
however, physical conditions (like density, temperature, chemical composition) could differ from one region to 
another. 
The study of NGC\,7538 cluster regions \citep[IRS\,4-6, IRS\,1-3, and IRS\,9, which follow an age sequence in descending 
order;][]{ojha04a}, being part of the same complex, should mitigate this problem. 
Here, using the deepest NIR data, we have obtained the KLF and the MF of the IRS\,1-3 region, while only the KLF of the 
IRS\,9 region has been discussed. 
The MF for the IRS\,1-3 region shows that it rises till the BD limit before turning over, which indicates lower 
temperature or denser gas distribution than for Orion nebula cluster \citep{hillenbrand00,muench02}, though comparison 
of LF and MF of different regions from literature always follow the caveat that the method used for their construction 
might be slightly different. Presence of dense molecular material has been attested by surveys in submillimetre ranges 
too \citep{sandell04, reid05}. Our molecular line analysis also reveals massive clumps which could fragment and lead to 
future stellar cluster formation. The 850\,$\mu$m emission \citep{chavarria14} shows the IRS\,1-3 region to be located  
on a junction of filaments, which could partially explain the active star formation going on. Confinement of most IR 
clusters at filament junctions has also been observed by \citet{schneider12} and obtained in the simulations of 
\citet{dale12}. The entire morphology of NGC\,7538 seems similar to the hub-filament structure proposed by \citet{myers09}, 
where the IRS\,1-3 cluster region forms the \textquoteleft hub\textquoteright\, to which various filaments merge. The 
molecular hydrogen column density \citep[which is $> 10^{23}\,cm^{-3}$;][]{fallscheer13} satisfies the condition which 
\citet{myers09} proposes for a \textquoteleft hub\textquoteright\,.

The IRS\,9 region, however, appears to be distinct and at one end of the filament which connects it 
with IRS\,1-3 \citep[this filament goes on to connect IRS\,4 too; see][]{sandell04}. IRS\,9 region has been found to be 
much younger than IRS\,1-3, owing to a lack of free-free radio emission as well as a low number of YSOs associated with 
the nebula. The \textit{Herschel} temperature maps \citep{fallscheer13} also show that IRS\,9 is much colder than 
IRS\,1-3 region. The decreasing age-sequence along the filament connecting the three cluster regions of IRS\,4-6, IRS\,1-3, 
and IRS\,9 alludes to the possibility that star formation could have been triggered along this filamentary structure. 

To gain a firm understanding of the NGC\,7538 region stellar population, as well as 
the evolution of the KLF and MF as one moves from relatively older to relatively younger regions 
in a star-forming region, we need to obtain the detailed LFs and MFs for the IRS\,4-6 region as well as IRS\,9 region. The  
advantage will be that all the sub-regions belong to the same region NGC\,7538, and should therefore differences 
arising due to the differences in physical conditions (leading to different temperature and density distributions, 
which leads to different Jeans masses and thus different LFs/MFs) need not be a concern. Future analysis of the BDs 
detected in the region, construction of MF from a spectroscopic sample, an analysis of core mass function of this 
region, and examination of spectral energy distributions of individual sources are also required for a better 
idea of the star formation going on.

\section{Conclusions} 
\label{section_conclusion}

We have carried out the deep NIR imaging survey of IRS\,1--3 ($J, H$, and $K$) and IRS\,9 
($H$ and $K$) sub-regions 
in the NGC\,7538 star-forming region with the highest spatial resolution so far. In addition, GMRT 
observations at 325, 610, and 1280\,MHz are used to examine the radio emission and physical characteristics. 
H$^{13}$CO$^{+}$ ($J$=1--0) molecular line emission from Nobeyama radio telescope is used to understand the morphology. 
Our main results are summarized as follows. 
\begin{enumerate}
\item 
Based on the NIR CC-D ($J-H/H-K$), 144 Class\,II-type, and 24 Class\,I-type YSOs were identified in the 
IRS\,1--3 region. Using the NIR CM-D ($K/H-K$), 145 sources were identified in the IRS\,1-3 region, and 96 in  
IRS\,9 region. 27 sources were found to have X-ray counterparts. 
\item 
In the IRS\,1--3 region, the red sources and the Class\,I sources are concentrated around the compact nebula associated with  
luminous IRS sources, and the H$^{13}$CO$^{+}$ ($J$=1--0) contours show a peak (column density n(H$^{13}$CO$^{+}$) 
$\sim 3.6 \times 10^{12}$\,cm$^{-2}$) to the south of these sources. Stellar surface density analysis reveals three 
clusterings in this region. 
The IRS\,9 region does not have any particular distribution of the YSOs, with the nebula hardly containing any sources. 
An H$^{13}$CO$^{+}$ ($J$=1--0) peak (n(H$^{13}$CO$^{+}$) $\sim 2.7 \times 10^{12}$\,cm$^{-2}$) lies to the east of the 
cluster around IRS\,9 source. 
The virial masses are approximately of the order of 1000\msun\, and 500\msun\, for 
the clumps in IRS\,1--3 and IRS\,9 regions, respectively. 
\item 
Radio emission shows a champagne flow morphology in the NGC\,7538 region. In low-resolution 325 and 610\,MHz maps, a compact 
\HII\, region is seen associated with IRS\,1--3 sources, with its spectral index calculated to be $0.87 \pm 0.11$, 
suggesting optically thickness. 
In very high resolution 1280\,MHz maps, the IRS\,1--3 compact \HII\, region is resolved into three 
separate peaks, one of which coincides with the known IRS\,2 source. 
\item 
KLFs were constructed for IRS\,1--3 region for two sets of sources : 
\textquoteleft F+T+P+red-sources\textquoteright\, and 
\textquoteleft T+P+Any source with X-ray detection\textquoteright\,. The rising part of the KLF has a 
$(d\log N/dm_{K})$ slope of $0.40 \pm 0.03$ for the first set, and $0.24 \pm 0.03$ for the second set ($K$ magnitude range 
12.5--16.5). The KLF slopes for the IRS\,9 region using sources with $K$ band only detections 
was calculated to be $0.23 \pm 0.04$ (13--19 mag).  
\item 
Theoretical mass-luminosity relation is used to obtain $\xi(\log M_*)$ (differential form of MF) and $N(\log M_*)$ for 
the IRS\,1--3 cluster region. Both $\xi(\log M_*)$ and $N(\log M_*)$ show a peak in the low mass regime as well as a peak 
in the intermediate mass regime. In low-mass regime, $\xi(\log M_*)$ extends upto BD limit before turn-off 
(0.08--0.1 M$_\odot$), while $N(\log M_*)$ peak is in 0.1--0.18 M$_\odot$ range. 
The slope $d\log N/d\log M_*$ (from $N(\log M_*)$ distribution) for the first and second YSO sets in the range 
0.1--1 M$_\odot$ are $0.54 \pm 0.05$ and $0.75 \pm 0.18$, respectively - much lower than the Salpeter value of 1.35. 
The MFs most closely resemble that of \citet{scalo86}. The star to BD ratio upper limit was calculated to be 10.2.  
\end{enumerate}

\section*{Acknowledgments}
We thank the anonymous referee for a critical reading of the manuscript and several 
useful comments and suggestions, which greatly improved the scientific content of the paper.
This research made use of data collected at Subaru Telescope, which is 
operated by the National Astronomical Observatory of Japan. We are 
grateful to the Subaru Telescope staff for their support. 
We thank the staff of GMRT managed by National Center for Radio Astrophysics 
of the Tata Institute of Fundamental Research (TIFR) for their assistance 
and support during observations. 
D.K.O. was supported by the National Astronomical Observatory of Japan (NAOJ), Mitaka,  
through a fellowship, during which part of this work was done. This 
research was partly supported by Grants-in-Aid for Scientific Research 
on Priority Areas, ``Development of Extra-Solar Planetary Science'', and 
is partly supported by Grants-in-Aid for Specially Promoted Research,  
from the Ministry of Education, Culture, Sports, Science and Technology 
of Japan (16077101, 16077204), and by JSPS (16340061). 
K.K.M., D.K.O., I.Z., and L.P. acknowledge support from DST-RFBR Project 
(P-142; 13-02-92697) under the auspices of which some part of this work was 
carried out. 
S.D. is supported by a Marie-Curie Intra European Fellowship under 
the European Community's Seventh Framework Program FP7/2007-2013 grant 
agreement no 627008. I.Z. and L.P. are also partly supported by the grant within 
the agreement of August 27, 2013 No. 02.B.49.21.0003 between The Ministry of education 
and science of the Russian Federation and Lobachevsky State University of Nizhni Novgorod.

\begin{table*}
\centering
\caption{90\% Completeness Limits for NIR bands}
\label{table_completeness}
\begin{tabular}{@{}cccc}
\hline 
 & $K$ & $H$ & $J$ \\
 & (mag) & (mag) & (mag) \\
\hline 
\multicolumn{4}{c}{IRS\,1-3 region} \\
\hline
Overall  &   18    &   19    &   20.2  \\
1        &   17    &   18.4  &   19.2  \\
2        &   18.5  &   19.4  &   20.2  \\
3        &   18    &   20    &   19.7  \\
\hline
\multicolumn{4}{c}{IRS\,9 region} \\
\hline
Overall  &   17.7  &  16.6  &   \\
1        &   19.2  &  16.6  &   \\
2        &   17.7  &  16.6  &   \\
\hline
\multicolumn{4}{c}{Sky region} \\
\hline
Whole    &   20 & 21 & 21.2 \\
\hline
\end{tabular}
\end{table*}

\begin{table*}
\centering 
\caption{Details of Radio Continuum Observations}
\label{table_RadioObservation} 
\begin{tabular}{@{}ccccc}
\hline 
 & 1280 MHz & 610 MHz & 325 MHz & VLA 4860\,MHz Archival Image \\
\hline 
Date of Obs. & 2004 January 25 & 2004 September 18 & 2004 July 03 & 2000 September 22 (BP0068) \\
Phase Center & $\alpha_{2000}$ = 23$^h$13$^m$44$^s$ & $\alpha_{2000}$ = 23$^h$13$^m$45.28$^s$ & $\alpha_{2000}$ = 23$^h$13$^m$45.28$^s$ &  \\
             & $\delta_{2000}$ = 61$^o$28$^{'}$44.24$^{''}$ & $\delta_{2000}$ = 61$^o$28$^{'}$09.07$^{''}$ &  $\delta_{2000}$ = 61$^o$28$^{'}$09.07$^{''}$ &  \\
Flux Calibrator & 3C48 & 3C48 & 3C48, 3C147 &  \\
Phase Calibrator & 2355+498 & 2350+646 & 2350+646 &  \\
Cont. Bandwidth & 16 MHz & 16 MHz & 16 MHz &  \\
Primary Beam & 26.2$^{'}$ & 43$^{'}$  & 81$^{'}$  &  \\
Resolution of maps & & & & \\
used for fitting & 11.5$^{''} \times$\,8.5$^{''}$ & 12.0$^{''} \times$\,8.1$^{''}$ & 12.3$^{''} \times$\,8.7$^{''}$ & 14.9$^{''} \times$\,11.5$^{''}$ \\
rms noise & 1.93\,mJy\,beam$^{-1}$ & 6.82\,mJy\,beam$^{-1}$ & 1.78\,mJy\,beam$^{-1}$ & 1.56\,mJy\,beam$^{-1}$ \\ 
Integrated flux density & & & & \\
for IRS\,1-3 region & 0.69\,Jy & 0.30\,Jy & 0.14\,Jy & 1.53\,Jy \\
\hline
\end{tabular}
\end{table*}

\begin{landscape}
\begin{table} \centering
\caption{X-ray Sources with NIR counterparts in the NIR FoV}
\label{table_XraySources} 
% \begin{tiny}
\begin{tabular}{@{}cccccccccccccc}
\hline 
\multicolumn{5}{c}{Source} & \multicolumn{6}{c}{Xspec (Using \textit{ACIS Extract})} & \multicolumn{3}{c}{Xphot} \\
\multicolumn{5}{c}{\hrulefill} & \multicolumn{6}{c}{\hrulefill} & \multicolumn{3}{c}{\hrulefill} \\
IAU Designation  &  RA    & Dec   & $C_{t,net}$  & $E_{median}$  & log N$_{H1}$ & kT$_1$  & log L$_{h1}$  & log L$_{t1}$ & log L$_{hc1}$ & log L$_{tc1}$ & 
log L$_{hc2}$   & log L$_{tc2}$ & log N$_{H2}$ \\
  &  (deg)    & (deg)   & (counts)  & (keV)  & (cm$^{-2}$) & (keV)  & (erg s$^{-1}$)  & (erg s$^{-1}$) & (erg s$^{-1}$) & (erg s$^{-1}$) & 
(erg s$^{-1}$)   & (erg s$^{-1}$) & (cm$^{-2}$) \\
(1)  &  (2)    & (3)   &  (4) & (5)  & (6)  &  (7) &  (8) & (9) & (10) & (11) & (12) & (13)  &  (14)  \\
% # CXOU J                        $\alpha$            $\delta$         $C_{t,net}$       $E_{median}$}     logNH1             kT1                Lh               Lt              Lhc              Ltc               Lhc                Ltc             logNH      
% #
\hline
  231336.03$+$612806.5     &       348.400166    &    61.468476     &      14.9      &         2.2     &   22.491     &      1.751     &      30.703     &     30.783     &      30.848     &     31.331     &      30.761     &       31.195     &     22.176  \\  
  231338.51$+$612847.0     &       348.410483    &     61.47973     &       4.9      &          3.     &   22.096     &      9.528     &      30.299     &     30.364     &      30.336     &     30.519     &         &          &      \\  
  231338.72$+$612836.9     &        348.41136    &    61.476943     &       8.9      &         2.5     &   22.699     &      1.591     &      30.581     &     30.627     &      30.815     &     31.348     &      30.802     &       31.063     &     22.301  \\  
  231339.47$+$612832.8     &        348.41448    &    61.475793     &      32.9      &         2.6     &   22.609     &      2.023     &      31.157     &     31.205     &      31.329     &     31.756     &      31.296     &       31.662     &     22.342  \\  
  231339.94$+$612756.8     &       348.416427    &    61.465791     &       5.9      &         4.1     &    22.78     &      5.005     &      30.623     &     30.635     &      30.791     &     31.028     &      31.044     &       31.499     &     23.041  \\  
  231340.28$+$612902.2     &       348.417839    &    61.483949     &       6.9      &         1.9     &   22.288     &      1.749     &      30.249     &     30.379     &      30.343     &     30.826     &          &           &     \\  
  231340.65$+$612847.0     &       348.419397    &    61.479739     &       6.9      &         2.3     &   22.109     &      6.095     &      30.443     &     30.517     &      30.483     &     30.699     &      30.429     &       30.924     &      22.23  \\  
  231340.78$+$612828.1     &       348.419949    &    61.474477     &      28.9      &         2.8     &   22.668     &      2.015     &      31.134     &     31.172     &      31.326     &     31.754     &       31.33     &       31.655     &     22.398  \\  
  231341.96$+$612742.6     &       348.424864    &     61.46184     &      15.9      &         3.5     &     22.7     &      8.762     &      31.058     &     31.071     &      31.187     &     31.375     &      31.225     &       31.649     &     22.708  \\  
  231342.29$+$612830.5     &       348.426229    &    61.475165     &       5.8      &         3.5     &   23.033     &      2.268     &      30.585     &      30.59     &      30.941     &     31.331     &      30.864     &       31.216     &     22.716  \\  
  231342.71$+$612848.7     &       348.427984    &    61.480202     &       4.9      &         3.1     &    22.16     &      9.526     &      30.349     &     30.406     &       30.39     &     30.574     &         &          &     \\  
  231343.97$+$612807.4     &        348.43321    &    61.468723     &      30.7      &         4.3     &   23.111     &      8.697     &      31.476     &     31.477     &      31.745     &     31.933     &      31.862     &       32.216     &     23.079  \\  
  231344.26$+$612809.8     &       348.434439    &    61.469415     &      11.7      &         3.8     &   22.377     &      9.528     &      30.771     &     30.807     &      30.837     &     31.021     &      31.254     &       31.659     &     22.875  \\  
  231345.02$+$612842.5     &       348.437602    &    61.478481     &       4.8      &          4.     &   22.637     &       9.52     &      30.487     &     30.503     &      30.598     &     30.782     &        &         &     \\  
  231345.05$+$612807.8     &       348.437735    &     61.46886     &      15.7      &         3.9     &   22.725     &      9.528     &      30.993     &     31.004     &      31.123     &     31.308     &      31.415     &       31.779     &     22.886  \\  
  231345.35$+$612807.9     &       348.438981    &    61.468886     &      25.7      &          4.     &   23.048     &      3.574     &      31.321     &     31.322     &       31.62     &     31.905     &      31.666     &       32.023     &     22.924  \\  
  231345.41$+$612825.4     &       348.439217    &    61.473744     &      35.8      &          4.     &   22.747     &      7.085     &      31.424     &     31.436     &       31.57     &     31.773     &      31.762     &       32.172     &     22.947  \\  
  231346.32$+$612752.8     &       348.443014    &    61.464692     &       4.8      &         3.8     &   23.171     &      1.908     &      30.542     &     30.544     &      31.032     &      31.48     &       &         &    \\  
  231347.06$+$612819.8     &       348.446089    &    61.472175     &       3.8      &         3.7     &   23.149     &      2.275     &       30.47     &     30.471     &      30.899     &     31.286     &        &        &     \\  
  231348.88$+$612822.6     &       348.453703    &     61.47296     &       6.9      &         3.7     &    22.42     &      9.528     &      30.977     &     31.011     &      31.052     &     31.234     &      31.332     &       31.737     &     22.778  \\  
  231352.30$+$612654.6     &       348.467923    &    61.448523     &       4.9      &         1.9     &   22.921     &      0.543     &      30.452     &     30.555     &      31.035     &     32.634     &       &        &    \\  
  231357.29$+$612815.5     &       348.488749    &    61.470973     &       5.9      &          5.     &   22.655     &      9.528     &      30.618     &     30.633     &      30.734     &     30.918     &      31.543     &       31.979     &     23.477  \\  
  231358.62$+$612819.3     &       348.494256    &    61.472035     &       5.9      &         3.7     &   22.583     &      9.528     &      30.585     &     30.605     &      30.686     &     30.869     &      30.915     &       31.315     &     22.785  \\  
  231400.86$+$612646.9     &       348.503605    &    61.446368     &       8.9      &         2.4     &   21.238     &      9.528     &      30.367     &     30.512     &      30.373     &     30.556     &      30.622     &       31.084     &     22.301  \\  
  231401.57$+$612752.9     &       348.506575    &    61.464702     &      87.9      &         2.5     &     22.4     &       4.23     &      31.775     &     31.823     &      31.856     &     32.115     &      31.841     &       32.173     &     22.255  \\  
  231401.78$+$612643.6     &       348.507435    &    61.445464     &       8.9      &         3.2     &   23.184     &      1.265     &      30.743     &     30.747     &      31.381     &     32.064     &       30.97     &       31.307     &     22.591  \\  
  231402.39$+$612720.1     &       348.509978    &    61.455607     &       6.9      &         3.6     &   22.411     &      9.528     &      30.611     &     30.644     &      30.681     &     30.865     &      30.955     &       31.389     &     22.771  \\  
\hline
\end{tabular}
% \end{tiny}

\begin{footnotesize}
Col.\ (1--3): IAU designation, RA and Dec. 
Col.\ (4--5): Net counts extracted in the total energy band (0.5--8~keV); Background-corrected median photon energy (total band). 
Col.\ (6--7): Column density and plasma temperature calculated using Xspec. 
Col.\ (8--9): Apparent hard and total band luminosity  
Col.\ (10--11): Intrinsic hard and total band luminosity  
Col.\ (12--14): Intrinsic hard band luminosity, total band luminosity, and the column density calculated using Xphot.   
\end{footnotesize}
\end{table}
\end{landscape}

\begin{table*}
\centering 
\caption{YSOs in IRS\,1-3 NIR FoV }
\label{table_YSOs_IRS13} 
\begin{tabular}{@{}cccccc}
\hline 
RA & Dec. & $J$ & $H$ & $K$ & YSO Classification, \\
(J2000) & (J2000) & (mag) & (mag) & (mag) & X-ray IAU Designation \\
\hline 
  348.438812    &   61.461308    &   21.014  $\pm$  0.024    &   17.398  $\pm$  0.001    &   15.335  $\pm$  0.034    &                           ClassIII     \\
  348.438904    &   61.486595    &   19.365  $\pm$  0.023    &   18.038  $\pm$  0.010    &   17.004  $\pm$  0.015    &                            ClassII     \\
  348.438995    &   61.468994    &                           &   17.973  $\pm$  0.010    &   14.514  $\pm$  0.026    &       $H-K>1$,231345.35$+$612807.9     \\
  348.439117    &   61.467808    &   20.258  $\pm$  0.030    &   15.619  $\pm$  0.010    &   12.824  $\pm$  0.095    &                            ClassII     \\
  348.439117    &   61.473797    &   16.536  $\pm$  0.034    &   14.370  $\pm$  0.020    &   13.220  $\pm$  0.023    &      ClassIII,231345.41$+$612825.4     \\
\hline
\end{tabular}

\begin{footnotesize}
The IAU designation is from Table \ref{table_XraySources}. 
Table~\ref{table_YSOs_IRS13} is available in its entirety in a machine-readable 
form in the online journal. A portion is shown here for guidance regarding its form and content.
\end{footnotesize}
\end{table*}

\begin{table*}
\centering 
\caption{YSOs in IRS\,9 NIR FoV }
\label{table_YSOs_IRS9} 
\begin{tabular}{@{}cccccc}
\hline 
RA & Dec. & $H$ & $K$ & X-ray IAU Designation \\
(J2000) & (J2000) & (mag) & (mag) &   \\
\hline
  348.46637   &   61.466164    &    16.278   $\pm$  0.012    &   15.167   $\pm$  0.005  &      \\ 
 348.466431   &   61.454922    &    18.088   $\pm$  0.022    &    15.91   $\pm$  0.009  &     \\ 
  348.46701   &   61.470901    &     17.93   $\pm$  0.013    &   16.225   $\pm$  0.013  &      \\ 
 348.467743   &   61.454628    &    19.445   $\pm$   0.03    &   17.492   $\pm$  0.017  &      \\ 
 348.467987   &   61.448483    &    15.983   $\pm$  0.009    &   14.191   $\pm$  0.004  &   231352.30$+$612654.6    \\ 
\hline
\end{tabular}

\begin{footnotesize}
The IAU designation is from Table \ref{table_XraySources}. 
Table~\ref{table_YSOs_IRS9} is available in its entirety in a machine-readable 
form in the online journal. A portion is shown here for guidance regarding its form and content.
\end{footnotesize}
\end{table*}

\clearpage

\clearpage

% \clearpage

\end{document}